%% file: main.tex
\pdfoutput=1
\documentclass[12pt,a4paper]{article}

\usepackage{ifthen} 
\newboolean{pdflatex}
\setboolean{pdflatex}{true} 

\newboolean{articletitles}
\setboolean{articletitles}{true} 

\newboolean{uprightparticles}
\setboolean{uprightparticles}{false} 

\def\paperauthors{LHCb collaboration} 
\def\paperasciititle{Amplitude analysis of the Xi^+_c pK^-pi^+ decay and Xi^+_c baryon polarization measurement in semileptonic beauty-hadron decays} 
\def\papertitle{Amplitude analysis of the $\PXi^+_c\to pK^-\pi^+$ decay and\\ $\PXi^+_c$ baryon polarization measurement\\ in semileptonic beauty-hadron decays} 
\def\paperkeywords{{High Energy Physics}, {LHCb}} 
\def\papercopyright{\the\year\ CERN for the benefit of the LHCb collaboration} 
\def\paperlicence{CC BY 4.0 licence}
\def\paperlicenceurl{https://creativecommons.org/licenses/by/4.0/}

\input{preamble}
\usepackage{longtable} 

\usepackage{amsmath}
\usepackage{booktabs,url}
\usepackage{braket}
\usepackage[utf8]{inputenc}
\usepackage{bm} 
\usepackage{bigints} 
\definecolor{darkgreen}{RGB}{0,102,0}
\usepackage{tikz}
\usetikzlibrary{arrows,matrix}

\include{thesis-symbols}

\graphicspath{{figs/}}

\def\Kstarbzz  {{\ensuremath{\Kbar{}^*_0}}\xspace}

\def\Kstarbtwo  {{\ensuremath{\Kbar{}^*_2}}\xspace}
\begin{document}

\renewcommand{\thefootnote}{\fnsymbol{footnote}}
\setcounter{footnote}{1}

\input{title-LHCb-PAPER}

\renewcommand{\thefootnote}{\arabic{footnote}}
\setcounter{footnote}{0}

\cleardoublepage


\pagestyle{plain} 
\setcounter{page}{1}
\pagenumbering{arabic}


\input{body}


\clearpage
\input{acknowledgements}

\input{appendix}


\addcontentsline{toc}{section}{References}
\bibliographystyle{LHCb}
\bibliography{main,standard,LHCb-PAPER,LHCb-CONF,LHCb-DP,LHCb-TDR}
 
\newpage
\input{Authorship_LHCb-PAPER-2024-034}

\end{document}

%% file: preamble.tex
\usepackage[top=1in, bottom=1.25in, left=1in, right=1in]{geometry}

\usepackage{microtype}
\usepackage{lineno}  
\usepackage{xspace} 
\usepackage{caption} 

\usepackage{graphicx}  
\usepackage{color}
\usepackage{colortbl}
\graphicspath{{./figs/}} 

\usepackage{amsmath} 
\usepackage{amssymb}
\usepackage{amsfonts}
\usepackage{upgreek} 

\newcommand*\patchAmsMathEnvironmentForLineno[1]{%
\expandafter\let\csname old#1\expandafter\endcsname\csname #1\endcsname
\expandafter\let\csname oldend#1\expandafter\endcsname\csname
end#1\endcsname
 \renewenvironment{#1}%
   {\linenomath\csname old#1\endcsname}%
   {\csname oldend#1\endcsname\endlinenomath}%
}
\newcommand*\patchBothAmsMathEnvironmentsForLineno[1]{%
  \patchAmsMathEnvironmentForLineno{#1}%
  \patchAmsMathEnvironmentForLineno{#1*}%
}
\AtBeginDocument{%
\patchBothAmsMathEnvironmentsForLineno{equation}%
\patchBothAmsMathEnvironmentsForLineno{align}%
\patchBothAmsMathEnvironmentsForLineno{flalign}%
\patchBothAmsMathEnvironmentsForLineno{alignat}%
\patchBothAmsMathEnvironmentsForLineno{gather}%
\patchBothAmsMathEnvironmentsForLineno{multline}%
\patchBothAmsMathEnvironmentsForLineno{eqnarray}%
}

\usepackage{hyperxmp}

\usepackage[pdftex,
            pdfauthor={\paperauthors},
            pdftitle={\paperasciititle},
            pdfkeywords={\paperkeywords},
            pdfcopyright={Copyright (C) \papercopyright},
            pdflicenseurl={\paperlicenceurl}]{hyperref}

\usepackage[colorinlistoftodos,textsize=scriptsize]{todonotes}

\usepackage[bottom,flushmargin,hang,multiple]{footmisc}

\usepackage[all]{hypcap} 

\input{lhcb-symbols-def} 

\usepackage{cite} 
\usepackage{mciteplus}

%% file: lhcb-symbols-def.tex
\usepackage{xspace} 
\usepackage{upgreek}

\def\lhcb   {\mbox{LHCb}\xspace}

\def\MagUp {\mbox{\em Mag\kern -0.05em Up}\xspace}

\ifthenelse{\boolean{uprightparticles}}%
{

 \def\Pmu         {\ensuremath{\upmu}\xspace}                 
 \def\Pnu         {\ensuremath{\upnu}\xspace}                 
                  
 \def\Ppi         {\ensuremath{\uppi}\xspace}

 \def\Ppsi        {\ensuremath{\uppsi}\xspace}

 \def\PDelta      {\ensuremath{\Delta}\xspace}                 
 \def\PXi         {\ensuremath{\Xi}\xspace}                 
 \def\PLambda     {\ensuremath{\Lambda}\xspace}                 
 \def\PSigma      {\ensuremath{\Sigma}\xspace}                 
 \def\POmega      {\ensuremath{\Omega}\xspace}                 
 \def\PUpsilon    {\ensuremath{\Upsilon}\xspace}
 \let\oldPi\Pi
 \def\PPi         {\ensuremath{\oldPi}\xspace}

 \def\PB      {\ensuremath{\mathrm{B}}\xspace}                 
                  
 \def\PD      {\ensuremath{\mathrm{D}}\xspace}

 \def\PJ      {\ensuremath{\mathrm{J}}\xspace}                 
 \def\PK      {\ensuremath{\mathrm{K}}\xspace}

 \def\Pb      {\ensuremath{\mathrm{b}}\xspace}                 
 \def\Pc      {\ensuremath{\mathrm{c}}\xspace}

 \def\Pi      {\ensuremath{\mathrm{i}}\xspace}

 \def\Pp      {\ensuremath{\mathrm{p}}\xspace}

 \def\Ps      {\ensuremath{\mathrm{s}}\xspace}

 \def\thebaroffset{0.0em}
}
{

 \def\Pmu         {\ensuremath{\mu}\xspace}                 
 \def\Pnu         {\ensuremath{\nu}\xspace}                 
                  
 \def\Ppi         {\ensuremath{\pi}\xspace}

 \def\Ppsi        {\ensuremath{\psi}\xspace}                 
                  
 \mathchardef\PDelta="7101
 \mathchardef\PXi="7104
 \mathchardef\PLambda="7103
 \mathchardef\PSigma="7106
 \mathchardef\POmega="710A
 \mathchardef\PUpsilon="7107
 \mathchardef\PPi="7105
                  
 \def\PB      {\ensuremath{B}\xspace}                 
                  
 \def\PD      {\ensuremath{D}\xspace}

 \def\PJ      {\ensuremath{J}\xspace}                 
 \def\PK      {\ensuremath{K}\xspace}

 \def\Pb      {\ensuremath{b}\xspace}                 
 \def\Pc      {\ensuremath{c}\xspace}

 \def\Pi      {\ensuremath{i}\xspace}

 \def\Pp      {\ensuremath{p}\xspace}

 \def\Ps      {\ensuremath{s}\xspace}

 \def\thebaroffset{0.18em}
}
\newcommand{\offsetoverline}[2][\thebaroffset]{\kern #1\overline{\kern -#1 #2}}%

\makeatletter
\ifcase \@ptsize \relax
  \newcommand{\miniscule}{\@setfontsize\miniscule{4}{5}}
\or
  \newcommand{\miniscule}{\@setfontsize\miniscule{5}{6}}
\or
  \newcommand{\miniscule}{\@setfontsize\miniscule{5}{6}}
\fi
\makeatother

\DeclareRobustCommand{\optbar}[1]{\shortstack{{\miniscule (\rule[.5ex]{1.25em}{.18mm})}
  \\ [-.7ex] $#1$}}

\def\mun        {{\ensuremath{\Pmu^-}}\xspace} 

\def\mumu       {{\ensuremath{\Pmu^+\Pmu^-}}\xspace}

\def\ellm       {{\ensuremath{\ell^-}}\xspace}

\def\neub       {{\ensuremath{\overline{\Pnu}}}\xspace}

\def\neumb      {{\ensuremath{\neub_\mu}}\xspace}

\def\neulb      {{\ensuremath{\neub_\ell}}\xspace}

\def\squark    {{\ensuremath{\Ps}}\xspace}

\def\cquark    {{\ensuremath{\Pc}}\xspace}

\def\bquark    {{\ensuremath{\Pb}}\xspace}

\def\pion   {{\ensuremath{\Ppi}}\xspace}

\def\pip    {{\ensuremath{\pion^+}}\xspace}
\def\pim    {{\ensuremath{\pion^-}}\xspace}

\def\kaon    {{\ensuremath{\PK}}\xspace}
\def\Kbar    {{\ensuremath{\offsetoverline{\PK}}}\xspace}

\def\KorKbar {\kern \thebaroffset\optbar{\kern -\thebaroffset \PK}{}\xspace}

\def\Kp      {{\ensuremath{\kaon^+}}\xspace}
\def\Km      {{\ensuremath{\kaon^-}}\xspace}

\def\KS      {{\ensuremath{\kaon^0_{\mathrm{S}}}}\xspace}

\def\Kstarb  {{\ensuremath{\Kbar{}^*}}\xspace}

\def\D       {{\ensuremath{\PD}}\xspace}

\def\DorDbar {\kern \thebaroffset\optbar{\kern -\thebaroffset \PD}\xspace}
\def\Dz      {{\ensuremath{\D^0}}\xspace}

\def\Dp      {{\ensuremath{\D^+}}\xspace}
\def\Dm      {{\ensuremath{\D^-}}\xspace}

\def\DpDm    {\ensuremath{\Dp {\kern -0.16em \Dm}}\xspace}

\def\Dstarp  {{\ensuremath{\D^{*+}}}\xspace}

\def\Ds      {{\ensuremath{\D^+_\squark}}\xspace}

\def\B       {{\ensuremath{\PB}}\xspace}

\def\BorBbar {\kern \thebaroffset\optbar{\kern -\thebaroffset \PB}\xspace}

\def\Bd      {{\ensuremath{\B^0}}\xspace}

\def\BdorBdbar {\kern \thebaroffset\optbar{\kern -\thebaroffset \Bd}\xspace}
\def\Bu      {{\ensuremath{\B^+}}\xspace}

\def\Bs      {{\ensuremath{\B^0_\squark}}\xspace}

\def\BsorBsbar {\kern \thebaroffset\optbar{\kern -\thebaroffset \Bs}\xspace}

\def\jpsi     {{\ensuremath{{\PJ\mskip -3mu/\mskip -2mu\Ppsi}}}\xspace}

\def\Y#1S{\ensuremath{\PUpsilon{(#1S)}}\xspace}

\def\proton      {{\ensuremath{\Pp}}\xspace}

\def\Deltares    {{\ensuremath{\PDelta}}\xspace}

\def\Lz          {{\ensuremath{\PLambda}}\xspace}

\def\LorLbar     {\kern \thebaroffset\optbar{\kern -\thebaroffset \PLambda}\xspace}

\def\Sigmares    {{\ensuremath{\PSigma}}\xspace}

\def\Xires       {{\ensuremath{\PXi}}\xspace}

\def\Lc          {{\ensuremath{\Lz^+_\cquark}}\xspace}

\def\Xicp        {{\ensuremath{\Xires^+_\cquark}}\xspace}

\def\Lb           {{\ensuremath{\Lz^0_\bquark}}\xspace}

\def\Xibz         {{\ensuremath{\Xires^0_\bquark}}\xspace}

\newcommand{\decay}[2]{\ensuremath{#1\!\to #2}\xspace} 

\def\to                 {\ensuremath{\rightarrow}\xspace}

\def\AT#1     {\ensuremath{A_{\mathrm{T}}^{#1}}\xspace}           

\def\C#1      {\ensuremath{\mathcal{C}_{#1}}\xspace}                       
\def\Cp#1     {\ensuremath{\mathcal{C}_{#1}^{'}}\xspace}                    
\def\Ceff#1   {\ensuremath{\mathcal{C}_{#1}^{\mathrm{(eff)}}}\xspace}        
\def\Cpeff#1  {\ensuremath{\mathcal{C}_{#1}^{'\mathrm{(eff)}}}\xspace}       
\def\Ope#1    {\ensuremath{\mathcal{O}_{#1}}\xspace}                       
\def\Opep#1   {\ensuremath{\mathcal{O}_{#1}^{'}}\xspace}                    

\newcommand{\nospaceunit}[1]{\ensuremath{\text{#1}}}       
\newcommand{\aunit}[1]{\ensuremath{\text{\,#1}}}       

\newcommand{\tev}{\aunit{Te\kern -0.1em V}\xspace}
\newcommand{\gev}{\aunit{Ge\kern -0.1em V}\xspace}
\newcommand{\mev}{\aunit{Me\kern -0.1em V}\xspace}
\newcommand{\kev}{\aunit{ke\kern -0.1em V}\xspace}
\newcommand{\ev}{\aunit{e\kern -0.1em V}\xspace}
 
\newcommand{\mevc}{\ensuremath{\aunit{Me\kern -0.1em V\!/}c}\xspace}
\newcommand{\gevc}{\ensuremath{\aunit{Ge\kern -0.1em V\!/}c}\xspace}
\newcommand{\mevcc}{\ensuremath{\aunit{Me\kern -0.1em V\!/}c^2}\xspace}
\newcommand{\gevcc}{\ensuremath{\aunit{Ge\kern -0.1em V\!/}c^2}\xspace}

\def\mum  {\ensuremath{\,\upmu\nospaceunit{m}}\xspace}

\def\fb   {\ensuremath{\aunit{fb}}\xspace}
\def\invfb   {\ensuremath{\fb^{-1}}\xspace}

\def\ps   {\ensuremath{\aunit{ps}}\xspace}

\newcommand{\chisq}{\ensuremath{\chi^2}\xspace}

\newcommand{\chisqip}{\ensuremath{\chi^2_{\text{IP}}}\xspace}

\def\gsim{{~\raise.15em\hbox{$>$}\kern-.85em
          \lower.35em\hbox{$\sim$}~}\xspace}
\def\lsim{{~\raise.15em\hbox{$<$}\kern-.85em
          \lower.35em\hbox{$\sim$}~}\xspace}

\newcommand{\Real}{\ensuremath{\mathcal{R}e}\xspace}
\newcommand{\Imag}{\ensuremath{\mathcal{I}m}\xspace}

\def\sPlot{\mbox{\em sPlot}\xspace}

\def\pt         {\ensuremath{p_{\mathrm{T}}}\xspace}

\def\ptot       {\ensuremath{p}\xspace}

\def\evtgen     {\mbox{\textsc{EvtGen}}\xspace}

\def\geant      {\mbox{\textsc{Geant4}}\xspace}

\def\photos     {\mbox{\textsc{Photos}}\xspace}

\def\pythia     {\mbox{\textsc{Pythia}}\xspace}

\def\root       {\mbox{\textsc{Root}}\xspace}

\def\tensorflow {\mbox{\textsc{TensorFlow}}\xspace}

\def\tell1  {TELL1\xspace}
\def\ukl1   {UKL1\xspace}

\newcommand{\ie}{\mbox{\itshape i.e.}\xspace}

\newcommand{\lhcborcid}[1]{\href{https://orcid.org/#1}{\hspace*{0.1em}\raisebox{-0.45ex}{\includegraphics[width=1em]{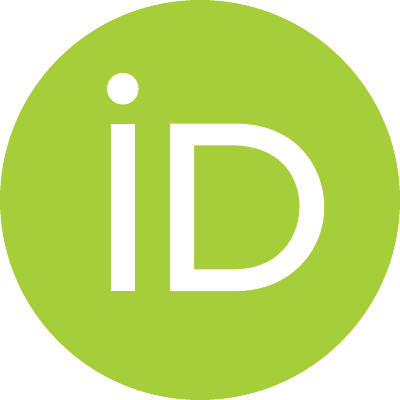}}}}


%% file: thesis-symbols.tex
\newcommand{\Lcpkpi}{\mbox{\decay{\Lc}{p\Km\pip}}\xspace}

\newcommand{\Dkpipi}{\mbox{\decay{\Dp}{\Km\pip\pip}}\xspace}
\newcommand{\Dskkpi}{\mbox{\decay{\Ds}{\Km\Kp\pip}}\xspace}
\newcommand{\Dkkpi}{\mbox{\decay{\Dp}{\Km\Kp\pip}}\xspace}
\newcommand{\Xicpkpi}{\mbox{\decay{\Xicp}{\proton\Km\pip}\xspace}}

\newcommand{\mqpk}{\ensuremath{m^2(p\Km)}\xspace}
\newcommand{\mqkpi}{\ensuremath{m^2(\Km\pip)}\xspace}


%% file: title-LHCb-PAPER.tex

\begin{titlepage}
\pagenumbering{roman}

\vspace*{-1.5cm}
\centerline{\large EUROPEAN ORGANIZATION FOR NUCLEAR RESEARCH (CERN)}
\vspace*{1.5cm}
\noindent
\begin{tabular*}{\linewidth}{lc@{\extracolsep{\fill}}r@{\extracolsep{0pt}}}
\ifthenelse{\boolean{pdflatex}}
{\vspace*{-1.5cm}\mbox{\!\!\!\includegraphics[width=.14\textwidth]{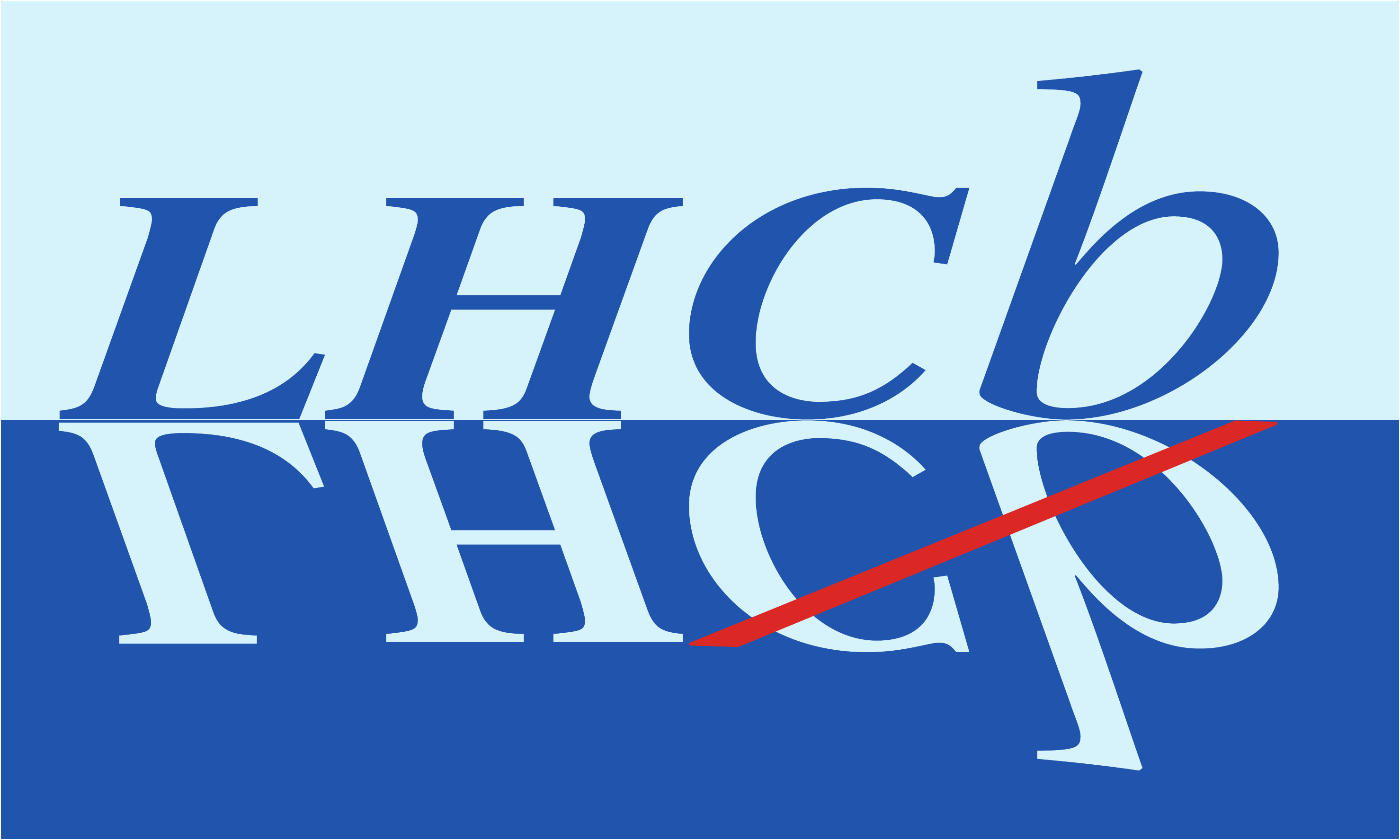}} & &}%
{\vspace*{-1.2cm}\mbox{\!\!\!\includegraphics[width=.12\textwidth]{figs/lhcb-logo.eps}} & &}%
\\
 & & CERN-EP-2024-320 \\  
 & & LHCb-PAPER-2024-034 \\  
 & & Nov 10, 2025 \\ 
 & & \\
\end{tabular*}

\vspace*{3.5cm} 

{\normalfont\bfseries\boldmath\huge
\begin{center}
  \papertitle 
\end{center}
}

\vspace*{2.0cm}

\begin{center}
\paperauthors\footnote{Authors are listed at the end of this paper.}
\end{center}

\vspace{\fill}

\begin{abstract}
  \noindent
An amplitude analysis of the $\PXi^+_c\to pK^-\pi^+$ decay together with a measurement of the $\PXi^+_c$ polarization vector in semileptonic beauty-hadron decays is presented. 
The analysis is performed using proton-proton collision data collected by the LHCb experiment, corresponding to an integrated luminosity of 9 ${\rm fb}^{-1}$. An amplitude model is developed and the resonance fractions as well as two- and three-body decay parameters are reported. A sizeable $\PXi^+_c$ polarization is found. A large sensitivity of the $\PXi^+_c\to pK^-\pi^+$ decay to the polarization is seen, making the amplitude model suitable for $\PXi^+_c$ polarization measurements in other systems.

\end{abstract}

\vspace*{2.0cm}

\begin{center}
  Published in
  Phys.~Rev.~D112 (2025) 092003
\end{center}

\vspace{\fill}

{\footnotesize 
\centerline{\copyright~\papercopyright. \href{\paperlicenceurl}{\paperlicence}.}}
\vspace*{2mm}

\end{titlepage}


\newpage
\setcounter{page}{2}
\mbox{~}
%
%
%
%

%% file: body.tex
\section{Introduction}
\label{sec:Introduction}

The Cabibbo-suppressed \Xicpkpi decay is of special interest for the study of \Xicp baryon properties due to its clean experimental signature, given by a displaced vertex composed by three quasi-stable charged hadrons, and a branching fraction of \mbox{$(0.62 \pm 0.30)\%$}~\cite{PDG2024}. Cabibbo-favored decays have larger branching fraction values, but they produce long-living hyperons characterized by lower reconstruction efficiencies. This makes the \Xicpkpi decay suitable for the study of \Xicp baryon properties at the \lhcb experiment, and even more at fixed-target experiments, where neutral hyperons have increased their flight distances due to the large Lorentz boost of the \Xicp baryons produced there. 

An amplitude analysis of the \Xicpkpi decay permits the simultaneous determination of the decay amplitudes characterizing intermediate resonant contributions and the measurement of the \Xicp polarization vector~\cite{Marangotto:2020ead}. Such an analysis has diverse applications, ranging from new physics searches to low-energy Quantum Chromodynamics (QCD) studies, as described in the following.

The helicity formalism is used to model each contribution to the \Xicpkpi decay, including intermediate-state polarization. The knowledge of the resonant structure is useful in searches for $C\!P$-symmetry violation in baryon decays~\cite{LHCb-PAPER-2016-059,LHCb-PAPER-2017-044,LHCb-PAPER-2018-001,LHCb-PAPER-2018-025,LHCb-PAPER-2018-044,LHCb-PAPER-2019-026,LHCb-PAPER-2019-028,LHCb-PAPER-2024-043}, as such asymmetries can be related to specific contributions or localized in phase space.
Parity violation is characterized by the decay-asymmetry $\alpha$ parameters associated to the two-body resonant contributions, which are determined from the helicity couplings. Parity violation is also studied for the entire three-body \Xicpkpi process, considering a quantity called average event information~\cite{Davier:1992nw}, which represents the sensitivity of the decay to the baryon polarization.
The \Xicpkpi decay amplitude model can be exploited in searches for physics contributions beyond the Standard Model  in $\decay{\Xibz}{\Xicp \ellm\neulb}$ decays~\cite{Zhang:2019xdm,Wang:2021ydv,Zhang:2019jax}. The spin analysis of the \Xicp baryon in angular analyses of \Xibz semileptonic decays increases both the number and sensitivity of measurable observables, notably that of the \Xicp longitudinal polarization.

The \Xicpkpi amplitude model can be employed to measure the polarization of the \Xicp baryon in multiple production processes. Polarization measurements of charm baryons are a fundamental probe of their spin structure and formation process via hadronization of heavy charm quarks. According to the heavy-quark effective theory, most of the $c$-quark polarization is expected to be retained by the charm baryon~\cite{Mannel:1991bs, Falk:1993rf, Galanti:2015pqa}. In the case of strong-force production, the baryon polarization is difficult to predict in the nonperturbative regime of QCD, thus its measurement discriminates among different low-energy QCD approaches. Moreover, polarization measurements enable  access to charm-baryon electric and magnetic dipole moments via spin precession~\cite{Botella:2016ksl,Bagli:2017foe,Fomin:2017ltw,Marangotto:2713231,Fomin:2019wuw,Mirarchi:2019vqi,Aiola:2020yam}.
The measurement of the \Xicp polarization, in comparison to that of the \Lc baryon, probes charm-baryon physics in the presence of strangeness.

Neither amplitude analyses of the \Xicpkpi decay, nor \Xicp polarization measurements, have been performed previously.
In this paper, an amplitude analysis of \Xicpkpi decays recorded by the \lhcb detector, including the measurement of the \Xicp polarization vector, is presented.
Throughout the paper, charge-conjugate states are implied. The analysis is based on a data sample of semileptonic decays of beauty hadrons produced in proton-proton (\proton\proton) collisions at center-of-mass energies of $7$, $8$ and $13\tev$, corresponding to an integrated luminosity of $9 \invfb$. The amplitude analysis closely follows the methods employed for the study of the \Lcpkpi decay at \lhcb~\cite{LHCb-PAPER-2022-002}.

This paper is organized as follows. The description of the \lhcb detector and the simulation sample employed is given in Sec.~\ref{sec:detector}. The selection of \Xicpkpi candidates and the invariant-mass fit used to determine signal and background yields are described in Sec.~\ref{sec:selection}. Amplitude and polarization fit frameworks and the development of the \Xicpkpi baseline amplitude model are described in Sec.~\ref{sec:amplitude_fit}. The evaluation of statistical and systematic uncertainties is covered by Sec.~\ref{sec:systematic}, along with the consistency checks performed. The results of the amplitude and polarization fits are reported in Sec.~\ref{sec:results}, and a brief summary of the analysis is provided in Sec.~\ref{sec:summary}.

\section{Detector and simulation}
\label{sec:detector}
The \lhcb detector~\cite{LHCb-DP-2008-001,LHCb-DP-2014-002} is a single-arm forward
spectrometer covering the \mbox{pseudorapidity} range $2<\eta <5$,
designed for the study of particles containing \bquark or \cquark
quarks. The detector includes a high-precision tracking system
consisting of a silicon-strip vertex detector surrounding the $pp$
interaction region, a large-area silicon-strip detector located
upstream of a dipole magnet with a bending power of about
$4{\mathrm{\,T\,m}}$, and three stations of silicon-strip detectors and straw
drift tubes placed downstream of the magnet.
The tracking system provides a measurement of the momentum, \ptot, of charged particles with
a relative uncertainty that varies from 0.5\% at low momentum to 1.0\% at 200\gev. Natural units with $c = 1$ are used throughout.
The minimum distance of a track to a primary $pp$ collision vertex (PV), the impact parameter, 
is measured with a resolution of $(15+29/\pt)\mum$,
where \pt is the component of the momentum transverse to the beam, in\,\gev.
Different types of charged hadrons are distinguished using information
from two ring-imaging Cherenkov detectors. 
Photons, electrons and hadrons are identified by a calorimeter system consisting of
scintillating-pad and preshower detectors, an electromagnetic
and a hadronic calorimeter. Muons are identified by a
system composed of alternating layers of iron and multiwire
proportional chambers.

The online event selection is performed by a trigger, 
which consists of a hardware stage, based on information from the calorimeter and muon
systems, followed by a software stage, which applies a full event
reconstruction.
At the hardware trigger stage, events are required to have at least one muon with high \pt.
  The software trigger requires a two-, three- or four-track
  secondary vertex with a significant displacement from any primary
  $pp$ interaction vertex. At least one charged particle
  must have a transverse momentum $\pt > 1.6\gev$ and be
  inconsistent with originating from a PV.
  A multivariate algorithm~\cite{BBDT,LHCb-PROC-2015-018} is used for
  the identification of secondary vertices consistent with the decay
  of a \bquark hadron.

The momentum scale is calibrated using samples of $\decay{\jpsi}{\mumu}$ 
and $\decay{\Bu}{\jpsi\Kp}$~decays collected concurrently
with the~data sample used for this analysis~\cite{LHCb-PAPER-2012-048,LHCb-PAPER-2013-011}.
The~relative accuracy of this
procedure is estimated to be $3 \times 10^{-4}$ using samples of other $\bquark$~hadrons, $\PUpsilon$~and
$\KS$~mesons.

Simulation is required to model the effects of the detector acceptance and the
  imposed selection requirements.
  In the simulation, $pp$ collisions are generated using
  \pythia~\cite{Sjostrand:2007gs,*Sjostrand:2006za}
  with a specific \lhcb configuration~\cite{LHCb-PROC-2010-056}.
  Decays of unstable particles
  are described by \evtgen~\cite{Lange:2001uf}, in which final-state
  radiation is generated using \photos~\cite{davidson2015photos}.
  The interaction of the generated particles with the detector, and its response,
  are implemented using the \geant
  toolkit~\cite{Allison:2006ve, *Agostinelli:2002hh} as described in
  Ref.~\cite{LHCb-PROC-2011-006}. 
  The underlying $pp$ interaction is reused multiple times, with an independently generated signal decay for each~\cite{LHCb-DP-2018-004}.

A sample of exclusive semileptonic \decay{\Xibz}{\Xicp\mun\neumb} decays is produced with the \Xicpkpi decay uniformly generated over the phase space. These simulated decays permit a suitable description of detector efficiency effects, despite the fact that other beauty-hadron decays could, in principle, contribute to the decay under study.
The particle identification (PID) response in the simulated samples is calibrated by sampling from data distributions of \decay{\Dstarp}{\Dz\pip}, \decay{\Dz}{\Km\pip} and \decay{\Lb}{\Lc \pim}, \Lcpkpi decays, considering their kinematics and the detector occupancy. An unbinned method is employed, where the probability density functions are modeled using kernel density estimation~\cite{Poluektov:2014rxa}.
Simulated distributions are validated on data, correcting observed differences using the gradient-boost weighting technique of the \mbox{\textsc{hep\_ml}}\xspace package\cite{Rogozhnikov:2016bdp}.

\section{Event selection and invariant-mass fit}
\label{sec:selection}

Data used in this analysis are split into a Run 1 dataset comprising \proton\proton collisions recorded at center-of-mass energies of $7$ and $8\tev$, and a Run 2 dataset recorded at $13\tev$ center-of-mass energy. They correspond to integrated luminosities of approximately 3 and \mbox{6\invfb}, respectively. The two datasets are treated in the same way but analyzed separately to consider possible differences due to the different data-taking conditions.

The \Xicp candidates are reconstructed from combinations of three charged hadron tracks forming a vertex separated from any PV. Semileptonic beauty-hadron decay candidates are then reconstructed inclusively requiring that the \Xicp candidate and a muon track originate from a common vertex. Small \chisq values are required for all track and vertex fits.
Each track is required to have a loose PID response; transverse momentum $\pt>250\mev$ for hadrons and $\pt>1\gev$ for muons; and momentum $p>2\gev$ for mesons, $p>8\gev$ for protons and $p>6\gev$ for muons.
Final-state particles are ensured to be well displaced from the interaction point by requiring a large track \chisqip with respect to any PV, where \chisqip is defined as the difference in the vertex fit \chisq of a given PV reconstructed with and without the particle under consideration. The PV with the smallest \chisqip is associated to the \Xicp candidate.

A set of selection criteria are imposed to reduce the combinatorial background, arising from random combinations of tracks. They require each track to be within the \lhcb detector pseudorapidity acceptance, $2<\eta<5$, and maximum decay times of $3$ and $10\ps$ for \Xicp and beauty-hadron candidates, respectively. Requirements are imposed also on proton and kaon PID probabilities, obtained from neural network classifiers, which combine information from the different subdetectors~\cite{LHCb-DP-2018-001}.

A boosted decision tree (BDT) classifier~\cite{Breiman} is used to further reduce the combinatorial background contamination. The classifier is trained on data passing the aforementioned selection criteria, with signal and background proxies obtained via an \sPlot technique~\cite{Pivk:2004ty}, separately for Run 1 and Run 2 datasets. It employs quantities related to decay topology, track fit quality, PID and the proton transverse momentum. A selection cut on the BDT classifier output is chosen optimising the figure of merit $N^2/(N+B)^{3/2}$, with $N$ and $B$ the signal and background yields. These are obtained in the $m(pK^-\pi^+)$ signal region within $\pm 15\mev$ of the known \Xicp mass~\cite{PDG2024}, covering approximately three times the mass-peak resolution.

The three-body charm-meson decays \Dkpipi, \Dskkpi and \Dkkpi  constitute the main physical background contributions.
They are identified by considering events with invariant mass reconstructed under the $\pip \Km \pip$ hypothesis within $\pm 15\mev$ of the known \Dp mass or invariant mass reconstructed under the $\Kp \Km\pip$ hypothesis within $\pm 15\mev$ of the known \Dp or \Ds masses. These events are then discarded only if the proton PID probability is lower than a given threshold. This requirement allows a more uniform efficiency over the \Xicpkpi decay phase-space distributions while maintaining an adequate background rejection.
Other physical background contributions are negligible after the selection process.
A background contribution originating from events in which the same positive track is reconstructed as a proton and a pion at the same time is removed by rejecting candidates in which the proton and pion momenta point in the same direction, \ie with relative angle difference below $10^{-3}$.

\begin{figure}
\centering
\includegraphics[width=0.49\textwidth]{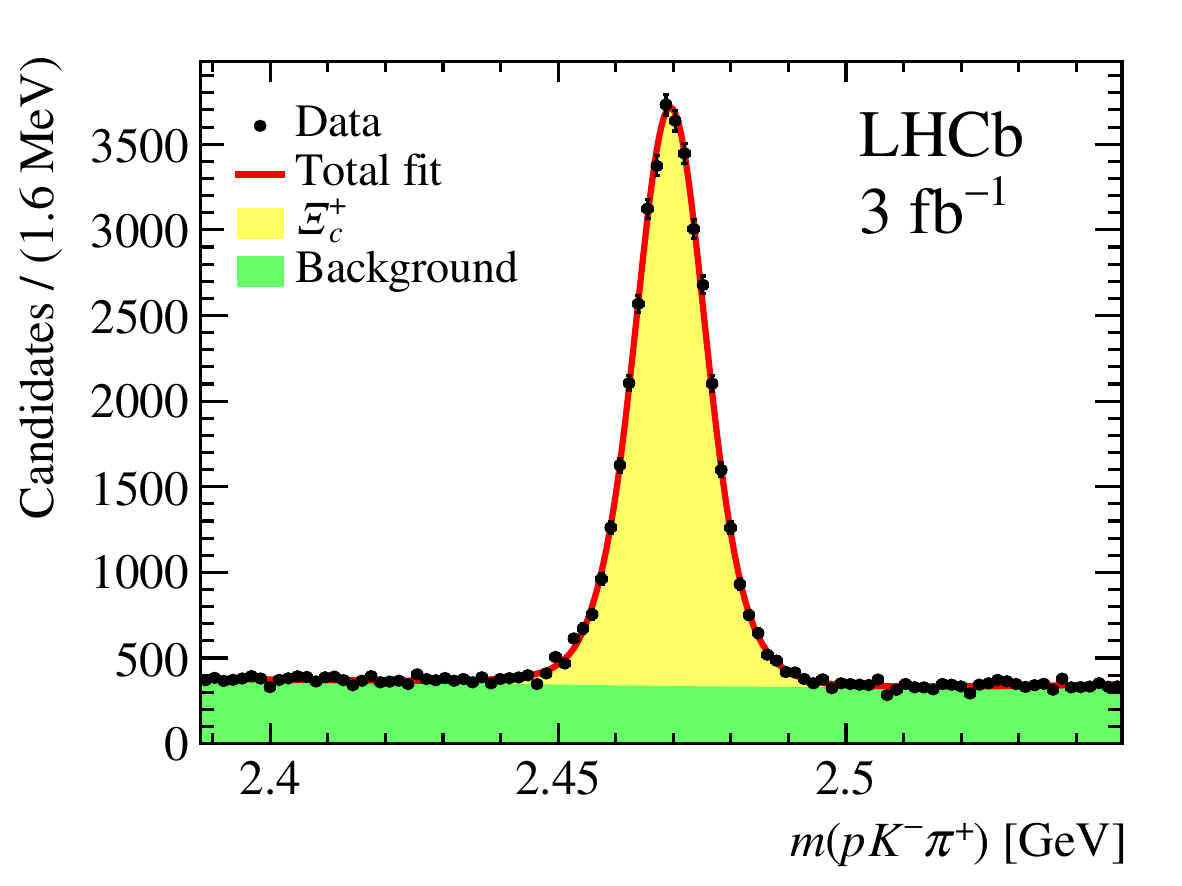}
\includegraphics[width=0.49\textwidth]{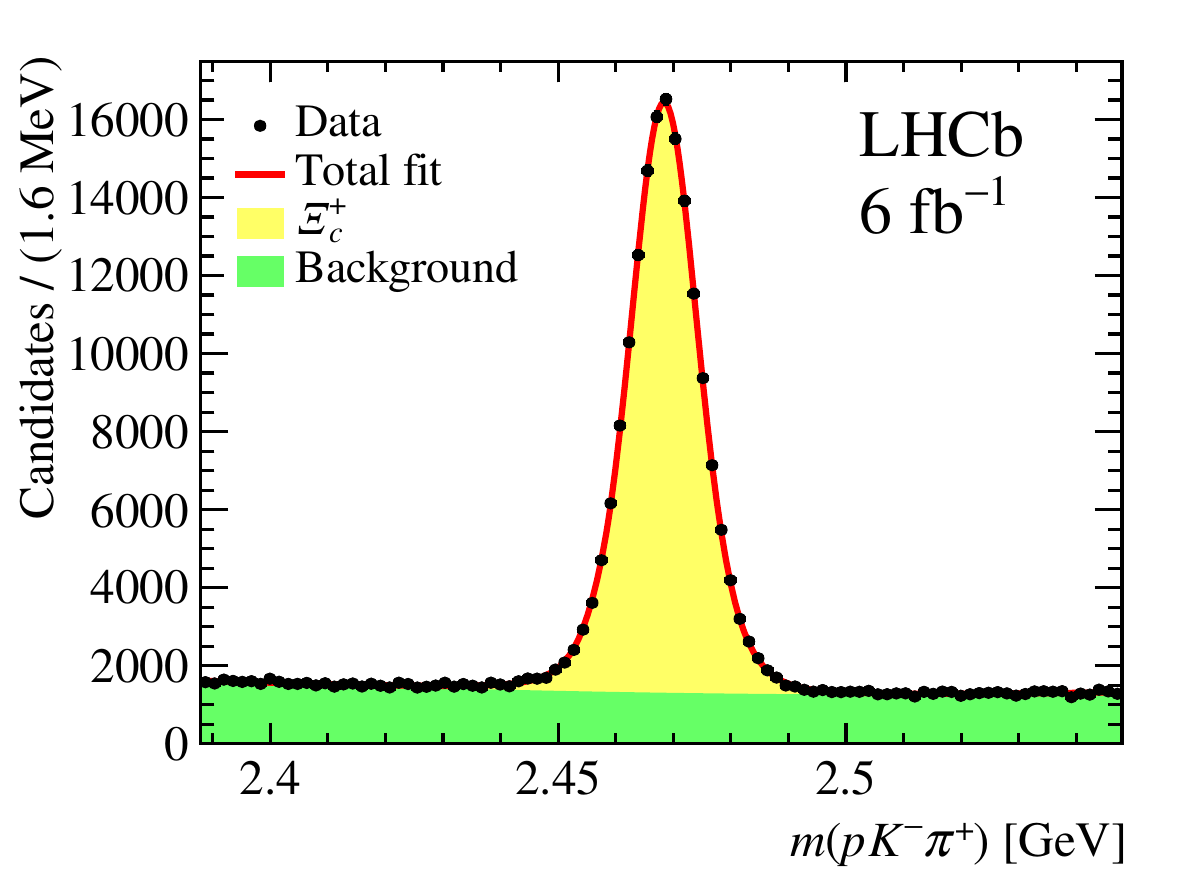}
\caption{Distributions of the $pK^-\pi^+$ invariant mass of (left) Run 1 and (right) Run 2 selected candidates.
The results from the fit described in the text are also shown.
\label{fig:fit}}
\end{figure}

The distribution of the invariant mass $m(\proton\Km\pip)$ of selected candidates from the Run~1 and Run 2 datasets is shown in Fig.~\ref{fig:fit}.
Extended unbinned maximum-likelihood fits are performed in the mass range within $\pm 80\mev$ of the known \Xicp mass, to determine \Xicpkpi and background yields after selection.
The \Xicpkpi signal shape except for the overall mean and width is fixed from simulation, while a polynomial function describes the combinatorial background contribution.

The signal $m(pK^-\pi^+)$ region chosen for the amplitude analysis is within $\pm 15\mev$ of the known \Xicp mass, containing $99.7\%$ of the signal candidates. Yields for the Run 1 and Run 2 data samples of 35265 and 151887 \Xicpkpi candidates are found, with background fractions in the signal region ($f_{\rm b}$) at $15.6\% $ and $14.4\%$, respectively.

\section{Amplitude and polarization fits}
\label{sec:amplitude_fit}
The amplitude model for the \Xicpkpi decay is written in the helicity formalism~\cite{JacobWick}, following the method and conventions of Ref.~\cite{Marangotto:2019ucc}. 
The \Xicp polarization is measured in the \Xicp rest frame in two different helicity systems defined by a boost along the \Xicp momentum from the approximate beauty-hadron rest frame ($\tilde{B}$) and from the laboratory frame (\textit{lab}).
The five variables describing the \Xicpkpi decay phase space are chosen to be
\begin{equation}
\Omega \equiv (\mqpk,\,\mqkpi,\,\cos\theta_\proton,\,\phi_\proton,\,\chi),
\label{eq:phase_space_vars}
\end{equation}
where \mqpk and \mqkpi are the squared invariant masses and $\theta_p,\phi_p$ and $\chi$ are the angles describing the decay orientation with respect to the \Xicp polarization system. The angles $\theta_p$ and $\phi_p$ are the polar and azimuthal angles of the proton momentum, while $\chi$ is the angle between the plane formed by the proton momentum and the \Xicp quantization axis, and the plane formed by the kaon and pion momenta, where momenta are expressed in the \Xicp rest frame. The phase-space variables are computed after constraining the mass of the $p\Km\pip$ candidate to the known \Xicp mass.
The detailed definition of the amplitude model, \Xicp helicity systems and phase-space variables is given in Ref.~\cite{LHCb-PAPER-2022-002}.

The free parameters of the amplitude model are determined by an unbinned maximum-likelihood fit to the five phase-space observables.
The detector efficiency effects on the \Xicpkpi phase space are included using calibrated simulation events passing event selection, while the background contribution is introduced using factorized Legendre-polynomial expansions, derived from data mass sidebands. Their distributions represent reliably the background contribution in the signal region since phase-space variables are mostly uncorrelated with the invariant mass $m(\proton\Km\pip)$. The maximum-likelihood fit is performed simultaneously on the separate Run 1 and Run 2 datasets, using specific simulation samples and background parametrizations for each dataset.

The amplitude fitting code is based on a version of the \mbox{\textsc{TensorFlowAnalysis}} package~\cite{TFA} adapted to three-body amplitudes in five-dimensional phase-space fits~\cite{Marangotto:2713231}. This package depends on the machine-learning framework \tensorflow~\cite{tensorflow2015-whitepaper} interfaced with \mbox{\textsc{Minuit}} minimization~\cite{James:1975dr} via the \root package~\cite{Brun:1997pa}.
The fit is performed using multiple gradient-descent minimization with different, randomized, initial values of the parameters to ensure that the best fit point in parameter space is reached.

The fit quality is measured by a \chisq test performed over a five-dimensional binning of the phase space; an adaptive binning technique is employed to guarantee a similar number of candidates in each bin. Probability values are obtained assuming a number of degrees of freedom equal to the number of bins employed minus the number of free parameters in the amplitude fit, including normalization.

The normalization of the model is determined by setting the complex coupling $\mathcal{H}^{\Kstarb(892)^0}_{1/2,0}$ to $(1,0)$, with the values of the other couplings expressed relative to this reference.
The baseline model is obtained from the amplitude fit in which the \Xicp polarization is expressed in the helicity system reached from the $\tilde{B}$ rest frame.
All the helicity couplings of the intermediate resonances are measured in the present analysis. Fit fractions and decay-asymmetry parameters are also determined for each two-body contribution.
In light of the application of the \Xicpkpi amplitude model as a \Xicp polarization analyzer, the sensitivity of the decay to the baryon polarization is measured by the normalized average event information, $\sqrt{3}S$, defined in Ref.~\cite{LHCb-PAPER-2022-002}. This quantity depends on the parity-violating part of the decay rate, and is inversely proportional to the variance of the polarization measurement.  The $\sqrt{3}S$ quantity is also measured for the nonzero-spin $K^*$ states in place of their decay-asymmetry parameters. 
\begin{figure}
\centering
\includegraphics[width=0.7\textwidth]{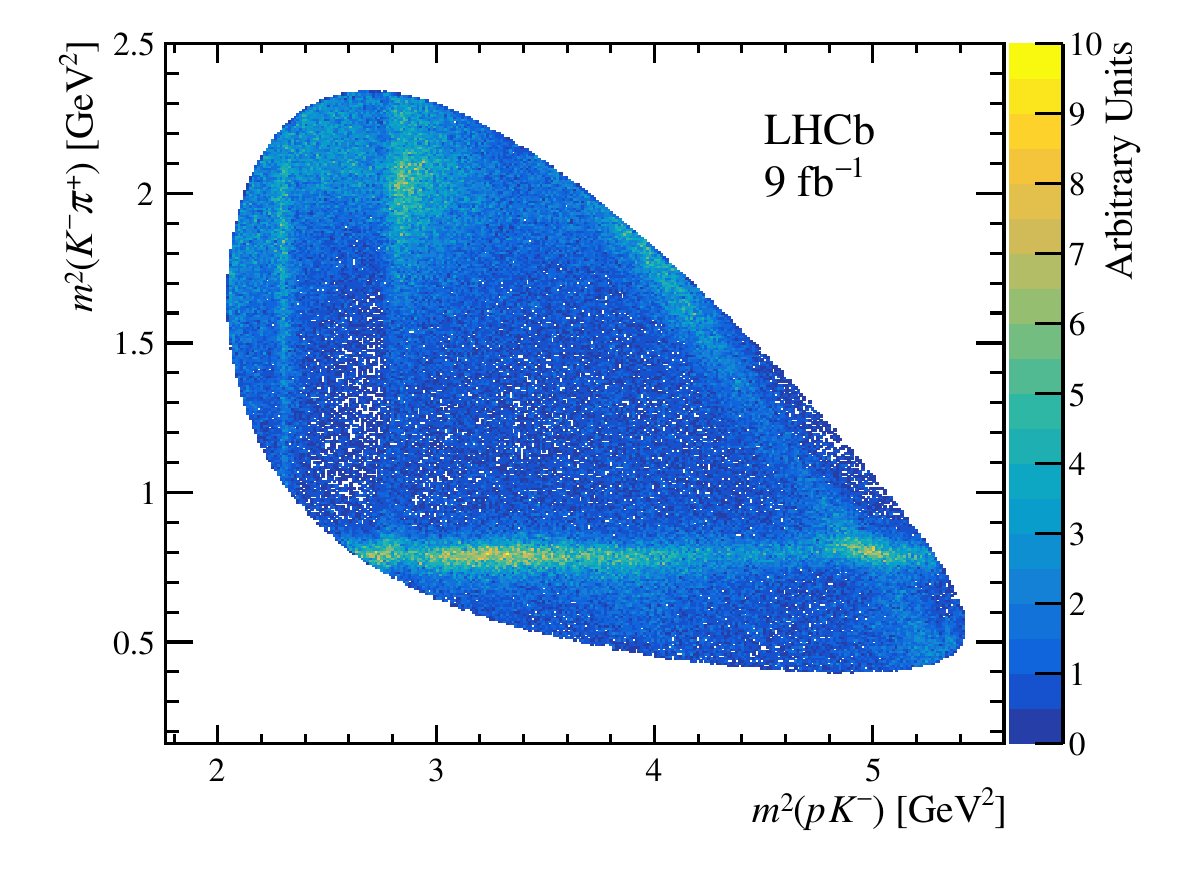}
\caption{Dalitz plot for the total sample of selected $\Xi^+_c\to p K^-\pi^+$ candidates. The sample consists of 187\,152 candidates with a purity of 85.4\%.\label{fig:data_dalitz_plot}}
\end{figure}

A Dalitz plot of the total reconstructed \Xicpkpi sample is presented in Fig.~\ref{fig:data_dalitz_plot}. The plot contains candidates from the signal region prior to efficiency correction, drawn in the \mqpk and \mqkpi squared invariant masses. It displays a rich structure with resonant contributions from all three possible pairs of final state particles: \Lz resonances are visible as vertical bands, \Kstarb as horizontal bands and $\Deltares^{++}$ as diagonal bands. The different intensity patterns can be explained by the spin of the resonance within each band, by interference patterns where bands overlap, or nonuniform detector efficiency near the boundaries of phase space.

The baseline amplitude model is built starting from the contributions visible in Fig.~\ref{fig:data_dalitz_plot} and adding resonant states according to those listed in Ref.~\cite{PDG2024}. Contributions which significantly improve the fit quality are added to the baseline model; those giving similar qualities are considered as alternative models for the evaluation of systematic uncertainties. The same criterion is employed for choosing among different descriptions of the same contribution.
\begin{table}
\centering
\caption{Resonant composition of the baseline \Xicpkpi amplitude model, with spin-parity $J^P$, and the Breit--Wigner mass and width parameters. \label{tab:nominal_model}}
\begin{tabular}{lcccc}
\toprule
Resonance & $J^P$ & Mass [\mev\!] & Width [\mev\!] & Reference\\
\midrule
$\Lz(1405)$ & $1/2^-$ & $1405.1$ & $50.5$ & \cite{PDG2024}\\
$\Lz(1520)$ & $3/2^-$ & $1518.467$ & $15.195$ & \cite{LHCb-PAPER-2022-002}\\
$\Lz(1600)$ & $1/2^+$ & $1600$ & $200$ & \cite{PDG2024}\\
$\Lz(1670)$ & $1/2^-$ & $1674.4$ & $27.2$ & \cite{Belle:2022cbs}\\
$\Lz(1690)$ & $3/2^-$ & $1690$ & $70$ & \cite{PDG2024}\\
$\Lz(1710)$ & $1/2^+$ & $1713$ & $180$ & \cite{PDG2024}\\
$\Lz(1800)$ & $1/2^-$ & $1800$ & $200$ & \cite{PDG2024}\\
$\Lz(1810)$ & $1/2^+$ & $1790$ & $110$ & \cite{PDG2024}\\
$\Lz(1820)$ & $5/2^+$ & $1820$ & $80$ & \cite{PDG2024}\\
$\Lz(1830)$ & $5/2^-$ & $1825$ & $90$ & \cite{PDG2024}\\
$\Lz(1890)$ & $3/2^+$ & $1890$ & $120$ & \cite{PDG2024}\\
$\Lz(2000)$ & $1/2^-$ & $1988.19$ & $179.26$ & \cite{LHCb-PAPER-2022-002}\\
\midrule
$\Kstarbzz(700)^0$ & $0^+$ & $845$ & $468$ & \cite{PDG2024}\\
$\Kstarb(892)^0$ & $1^-$ & $895.5$ & $47.3$ & \cite{PDG2024}\\
$\Kstarbzz(1430)^0$ & $0^+$ & $1425$ & $270$ & \cite{PDG2024}\\
$\Kstarbtwo(1430)^0$ & $2^+$ & $1432.4$ & $109$ & \cite{PDG2024}\\
\midrule
$\Deltares(1232)^{++}$ & $3/2^+$ & $1232$ & $117$ & \cite{PDG2024}\\
$\Deltares(1600)^{++}$ & $3/2^+$ & $1570$ & $250$ & \cite{PDG2024}\\
$\Deltares(1620)^{++}$ & $1/2^-$ & $1610$ & $130$ & \cite{PDG2024}\\
$\Deltares(1700)^{++}$ & $3/2^-$ & $1665$ & $250$ & \cite{PDG2024}\\
\bottomrule
\end{tabular}
\end{table}
The resonances included in the baseline model are listed in Table~\ref{tab:nominal_model}, where their invariant-mass dependence (lineshape) follows that of Ref.~\cite{LHCb-PAPER-2022-002}. Resonances are parametrized by default with relativistic Breit--Wigner functions multiplied by orbital angular-momentum-suppression terms.
The $\Lz(1405)$ and $\Lz(1670)$ resonances are parametrized by a Flatt\'{e} lineshape~\cite{Flatte:1976xu}.
Spin-zero $\Km\pip$ contributions, \ie $\Kstarbzz(700)^0$ and $\Kstarbzz(1430)^0$ states, are effectively described by a simplified version of the parametrization proposed in Ref.~\cite{Bugg:2005xx}.
It consists of a Breit--Wigner lineshape in which the mass-dependent width is given by
\begin{equation}
\Gamma(m) = \frac{m^2 - s_A}{m^2_0 - s_A} \Gamma_0 e^{-\gamma m^2},
\end{equation}
which features a singularity (Adler zero) at $s_A = m^2_K - 0.5 m^2_\pi$ and an exponential form factor on the $K\pi$ width driven by the parameter $\gamma$. The $\gamma$ parameter is determined separately for each contribution in the amplitude fit, with mass $m_0$ and width $\Gamma_0$ Breit--Wigner parameters taken from Ref.~\cite{PDG2024}.
The effect of this choice is taken into account in the computation of the model systematic uncertainty, described later, by releasing mass and width parameters in the fit, by varying the PDG values for the masses and widths by one standard deviation, and by considering relativistic Breit--Wigner functions as an alternative lineshape.
All resonance parameters are fixed to the mean values reported in Ref.~\cite{PDG2024} except for the $\Lz(1520)$ and $\Lz(2000)$ states, which are fixed to the central values determined by the \Lcpkpi amplitude analysis~\cite{LHCb-PAPER-2022-002}, and for the $\Lz(1670)$ resonance, which are fixed to the values determined by the Belle collaboration~\cite{Belle:2022cbs}. While no clear contribution of $\Sigmares^0$ states is found, their possible presence is considered among alternative models. The phase-space distributions are fully described by resonant contributions, with no need for additional nonresonant terms.

The \Xicp polarization in the $\tilde{B}$ system is determined directly by the amplitude fit, while that in the \textit{lab} system is measured via a separate fit in which the amplitude-model parameters are fixed to those determined in the other polarization system. This strategy, different with respect to that employed for the \Lcpkpi analysis~\cite{LHCb-PAPER-2022-002}, is motivated by the higher level of background contributions affecting the \Xicpkpi decay channel. In particular, charm-meson decays produce peaking structures in the \textit{lab} system $\cos\theta_p$ distribution. The effect on \textit{lab} system polarization due to the removal of this background source, which distorts the $\cos\theta_p$ distribution, is corrected for by exploiting both the simulation sample and the amplitude model determined in the $\tilde{B}$ frame.

\section{Uncertainties and consistency checks}
\label{sec:systematic}

Statistical and systematic uncertainties are computed for amplitude-model parameters, polarization components, fit fractions, decay-asymmetry parameters and $S^2$ quantities.
Statistical uncertainties are obtained by fitting the baseline model to 1000 pseudoexperiments sampled from fit results.
For each pseudoexperiment, the simulation sample used to compute the model normalization is also regenerated to account for its finite size. Two separate studies are performed for the amplitude fit, with polarization measured in the $\tilde{B}$ system, and for the \textit{lab} system polarization-only fit.
Statistical uncertainties for each parameter are determined as the standard deviation
of the distribution of results from the pseudoexperiments, reported along with the final results in Sec.~\ref{sec:results}.

Different sources of systematic uncertainties are considered. These are grouped into contributions coming from the model choice, the background determination, kinematics of the decay, PID, and the fit bias. For the \textit{lab} polarization fit, an additional contribution for the correction related to the removal of charm-meson decays is considered. The model contribution is quoted separately from the other experimental systematic contributions, which are combined in the final results.

The systematic uncertainty associated to the amplitude model choice is estimated by determining the measured parameters employing alternative models with fit quality similar to the baseline fit, which is the best representation of the resonant structure of the \Xicpkpi decay. Alternative models cover different types of baseline fit modifications. They include allowing the Breit--Wigner parameters of resonances with sizeable uncertainties to vary freely. Particular attention is given to the spin-zero $\Km\pip$ resonances, whose parameters are either floated or shifted by one standard deviation according to the PDG uncertainty intervals.
Alternative scenarios also explore the removal of the $\Lz(1800)$ and $\Lz(1890)$ states, which exhibit limited statistical significance in the fit. Their exclusion has no observable impact on the remaining resonant structure, although a significant change in the likelihood is observed when both are removed simultaneously.
The addition of contributions from the $\Lz(2100)$, $\Lz(2110)$, $\Sigmares(1670)^0$ and $\Sigmares(1775)^0$ states in the $\proton\Km$ decay channel is also investigated. Further variations include using relativistic Breit--Wigner functions as alternative lineshapes for spin-zero $\Km\pip$ resonances, and altering the orbital angular-momentum-suppression factor in the Breit--Wigner formulation. Each modification to the baseline amplitude model is applied independently. The maximum absolute deviation of the fitted parameters across all alternative models, relative to the baseline result, is assigned as the corresponding systematic uncertainty.

The uncertainty associated to the background description includes the uncertainty on the background fraction $f_{\rm b}$, estimated using an alternative model for the $m(p\Km\pip)$ mass shape, and that on the Legendre-polynomial parametrization, estimated varying both the background sample employed for its determination and the factorization of phase-space variables.
The uncertainties related to decay kinematics and PID corrections applied to the simulation are estimated separately varying the calibration samples employed and the functional form of the corrections. For the correction related to the removal of charm-meson decays in the \textit{lab} polarization fit, a different amplitude model is employed. A possible bias in the determination of fit parameters is considered by assigning the mean deviation of 1000 pseudoexperiments from the baseline results as a systematic uncertainty. 
Systematic uncertainties separated for each contribution are reported in  Table~\ref{tab:syst_pol} for polarization components, and in Appendix~\ref{sec:syst_summary} for fit parameters, fit fractions and decay asymmetries.
It is worth noting that the model uncertainty on polarization components is small compared to that associated to amplitude model parameters.

\begin{table}
\centering
\begin{small}
\caption{Systematic uncertainty contributions on polarization components in percentage. Total* includes all contributions except for the choice of amplitude model. Veto refers to the contribution connected to the removal of charm-meson decays. \label{tab:syst_pol}}
\begin{tabular}{lccccccc}
\toprule
Parameter & Model & Total* & Background & Kinematics & PID & Veto & Fit Bias \\
\midrule
$P_x$ ($\tilde{B}$) & 0.011\phantom{0} & 0.009\phantom{0} & 0.007\phantom{0} & \phantom{$<$}0.004\phantom{0} & \phantom{$<$}0.001\phantom{0} & & $<$0.001\phantom{0} \\
$P_y$ ($\tilde{B}$) & 0.0023 & 0.0035 & 0.0019 & \phantom{$<$}0.0023 & \phantom{$<$}0.0018 & & \phantom{$<$}0.0001 \\
$P_z$ ($\tilde{B}$) & 0.028\phantom{0} & 0.015\phantom{0} & 0.015\phantom{0} & $<$0.001\phantom{0} & $<$0.001\phantom{0} & & \phantom{$<$}0.002\phantom{0} \\
$P$ ($\tilde{B}$) & 0.030\phantom{0} & 0.016\phantom{0} & 0.016\phantom{0} & \phantom{$<$}0.002\phantom{0} & $<$0.001\phantom{0} & & \phantom{$<$}0.002\phantom{0} \\
\midrule 
$P_x$ (\textit{lab}) & 0.026\phantom{0} & 0.010\phantom{0} & 0.009\phantom{0} & $<$0.001\phantom{0} & \phantom{$<$}0.003\phantom{0} & 0.002\phantom{0} & $<$0.001\phantom{0} \\
$P_y$ (\textit{lab}) & 0.0019 & 0.0023 & 0.0002 & \phantom{$<$}0.0006 & \phantom{$<$}0.0015 & 0.0017 & \phantom{$<$}0.0001 \\
$P_z$ (\textit{lab}) & 0.014\phantom{0} & 0.020\phantom{0} & 0.007\phantom{0} & \phantom{$<$}0.007\phantom{0} & \phantom{$<$}0.014\phantom{0} & 0.010\phantom{0} & $<$0.001\phantom{0} \\
$P$ (\textit{lab}) & 0.030\phantom{0} & 0.012\phantom{0} & 0.010\phantom{0} & \phantom{$<$}0.003\phantom{0} & \phantom{$<$}0.003\phantom{0} & 0.006\phantom{0} & $<$0.001\phantom{0} \\
\bottomrule
\end{tabular}
\end{small}
\end{table}

The stability of the baseline amplitude model is checked by repeating the fit splitting the data set for different \Xicp charges, different data-taking periods and measuring the \Xicp polarization in the \textit{lab} polarization system. All amplitude models obtained are compatible with the baseline within uncertainties.

\section{Results}
\label{sec:results}

\begin{figure}
\centering
\includegraphics[width=\textwidth]{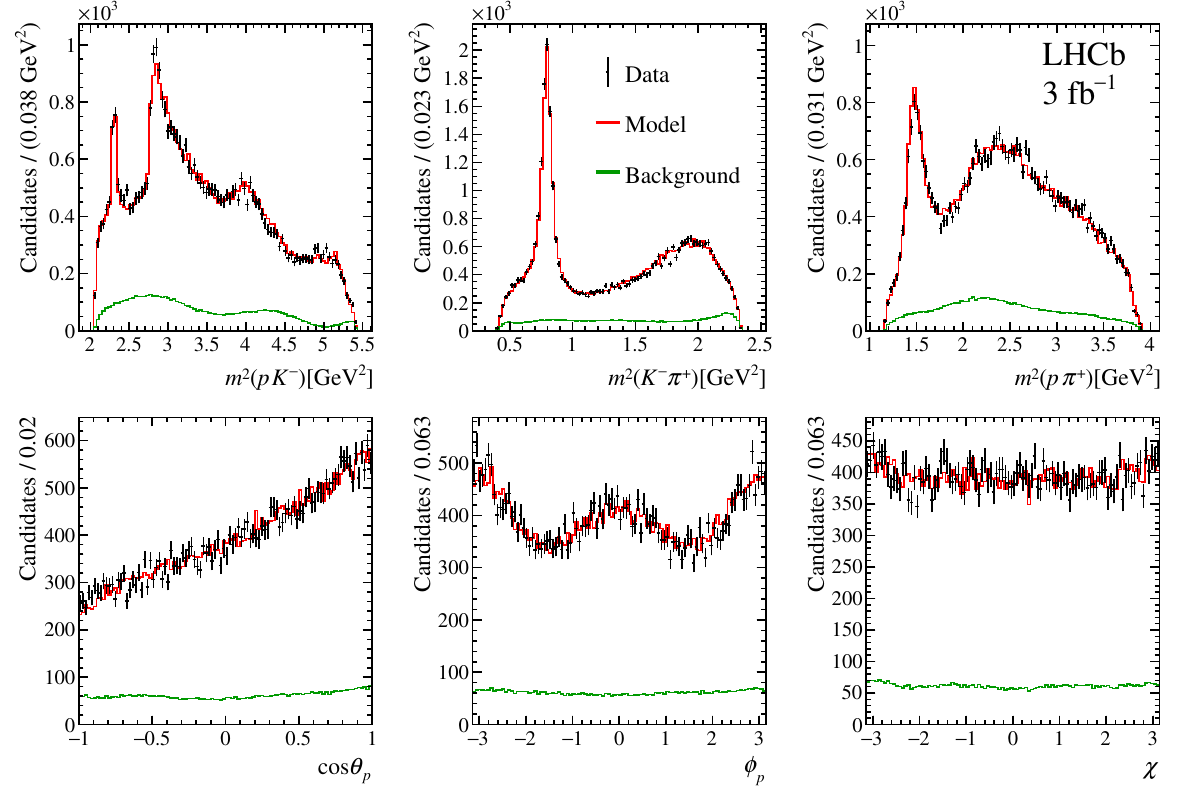}
\caption{Distributions for selected candidates in Run 1 together with amplitude fit projections in the $\tilde{B}$ system for (top row) invariant-mass-squared projections; and (bottom row) decay-orientation angle projections. \label{fig:data_boost_run1}}
\end{figure}

\begin{figure}
\centering
\includegraphics[width=\textwidth]{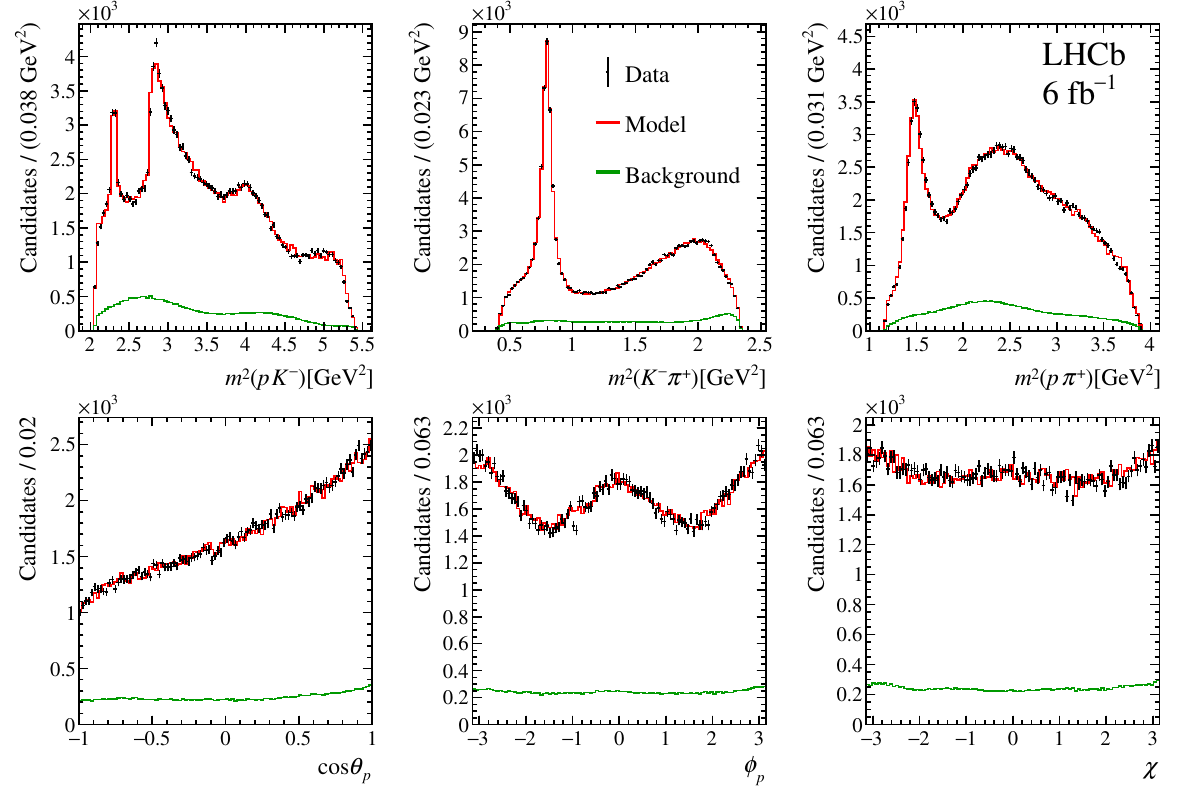}
\caption{Distributions for selected candidates in Run 2 together with amplitude fit projections in the $\tilde{B}$ system for (top row) invariant-mass-squared projections; and (bottom row) decay-orientation angle projections. \label{fig:data_boost}}
\end{figure}

\begin{figure}
\centering
\includegraphics[width=\textwidth]{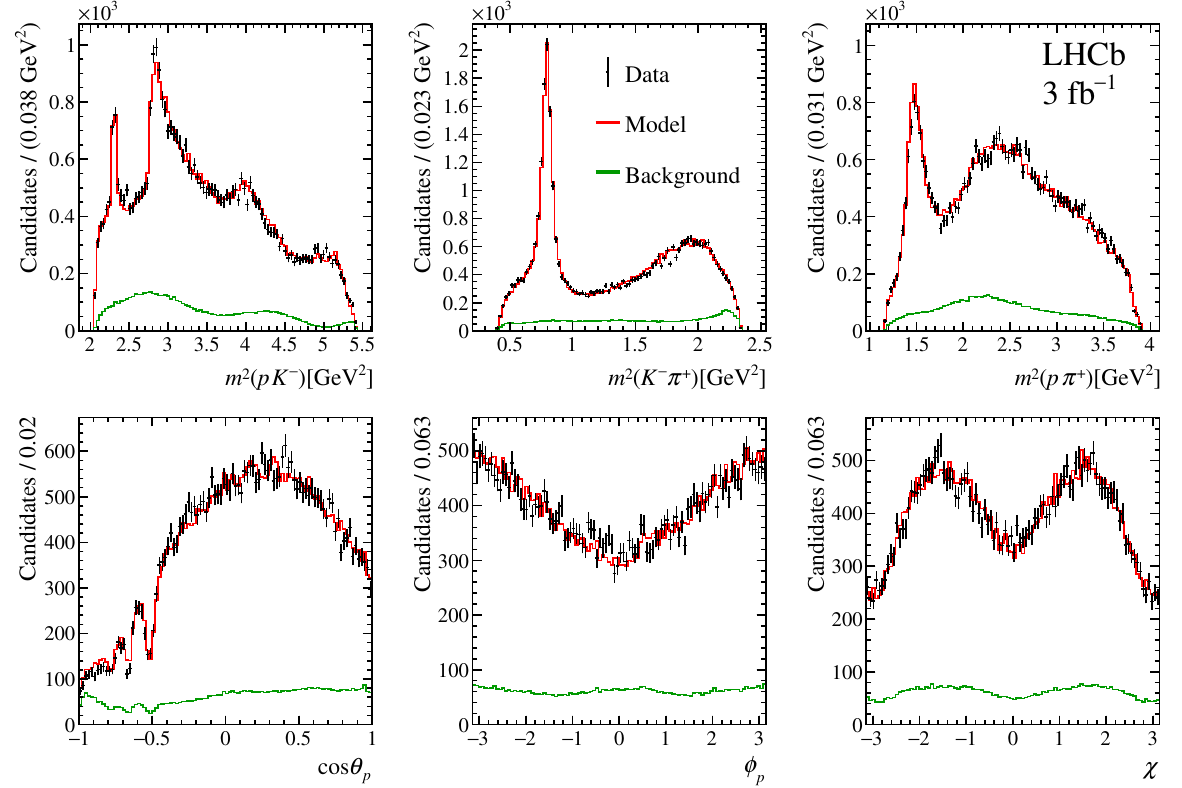}
\caption{Distributions for selected candidates in Run 1 together with polarization fit projections in the \textit{lab} system for (top row) invariant-mass-squared projections; and (bottom row) decay-orientation angle projections. \label{fig:data_lab_run1}}
\end{figure}

\begin{figure}
\centering
\includegraphics[width=\textwidth]{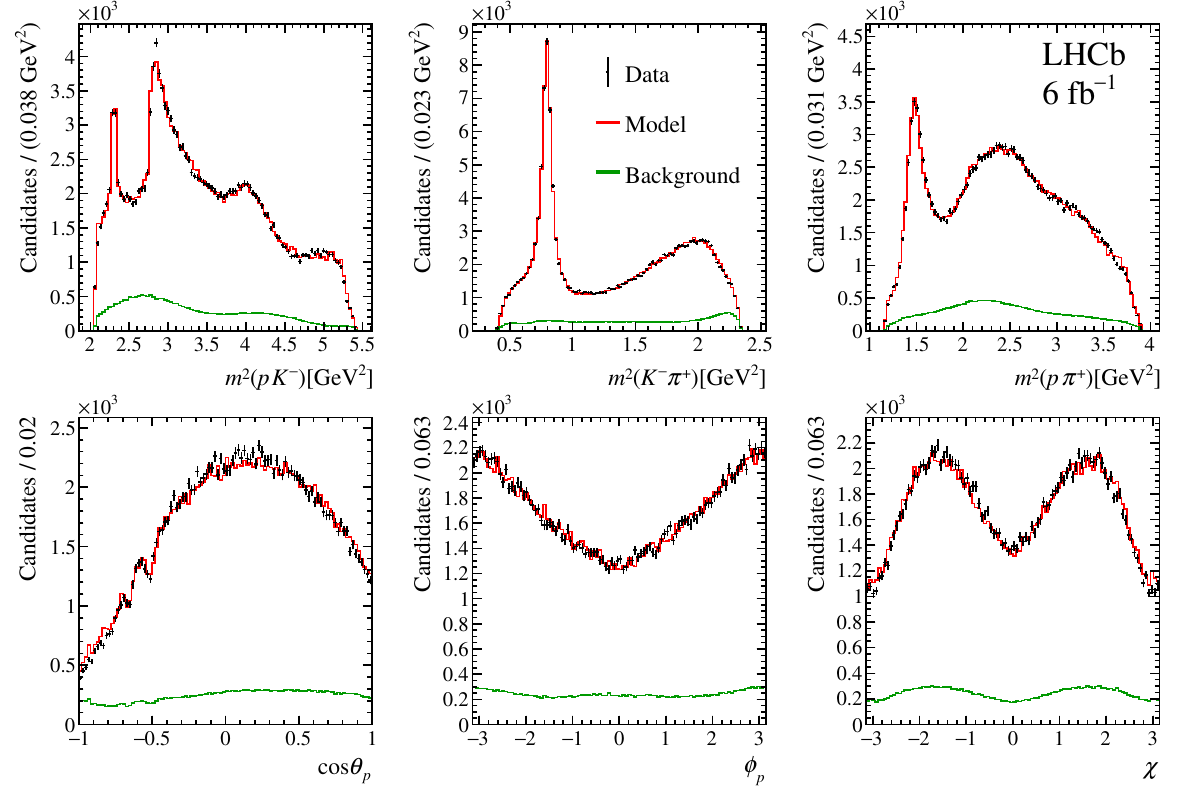}
\caption{Distributions for selected candidates in Run 2 together with polarization fit projections in the \textit{lab} system for (top row) invariant-mass-squared projections; and (bottom row) decay-orientation angle projections.\label{fig:data_lab}}
\end{figure}

The comparison between \Xicpkpi data and the baseline amplitude fit projections is displayed in Figs.~\ref{fig:data_boost_run1} and~\ref{fig:data_boost}, for Run 1 and Run 2 datasets, respectively. The comparison between \Xicpkpi data and \textit{lab}-system polarization fit projections is shown in Figs.~\ref{fig:data_lab_run1} and~\ref{fig:data_lab}. The amplitude-model distributions are obtained from the \Xicpkpi simulation sample which reproduces detector efficiency effects.

The polarization components in the $\tilde{B}$ and \textit{lab} systems are reported in Table~\ref{tab:results_final_pol}. The full \Xicp polarization vector is measured for the first time, with absolute uncertainties of order $1\%$ on each component. A large polarization is measured in both \Xicp helicity frames considered. In the $\tilde{B}$ system it has a magnitude $P \approx 66\%$, with a dominating negative longitudinal component $P_z \approx -62\%$ and a smaller positive transverse component $P_x \approx 22\%$.
In the \textit{lab} system it has a magnitude $P \approx 63\%$, with a dominating positive transverse component $P_x \approx 57\%$ and a smaller negative longitudinal component $P_z \approx -26\%$.
The normal polarization $P_y$, sensitive to time-reversal violation effects and final-state interactions~\cite{LHCb-PAPER-2022-002}, is compatible with zero at the $1\%$ level, for both polarization systems considered.
The leading uncertainty for longitudinal and transverse polarization in both systems is systematic,
while for normal polarization is statistical.

The \Xicp polarization shares the features observed for the \Lc baryon~\cite{LHCb-PAPER-2022-002} except for some slight differences in the measured values.
The \Xicp polarization direction is more transverse and less longitudinal than that of the \Lc one in the $\tilde{B}$ frame, more longitudinal and less transverse in the \textit{lab} frame, with a reduced magnitude in both polarization systems. Differences between \Xicp and \Lc polarization components have limited significance, being comparable in size with the uncertainties on the polarization components.

The measured parameters of the baseline amplitude model are reported in Tables~\ref{tab:results_final_Llow}-\ref{tab:results_final_D}, fit fractions for each resonant contribution in Table~\ref{tab:fitfractions_final}, $\sqrt{3}S$ and two-body decay-asymmetry parameters in Table~\ref{tab:alphapars_final}.
The amplitude analysis is sensitive to all the parameters describing the \Xicpkpi amplitude model, due to the sizeable \Xicp polarization and interference effects among different decay chains~\cite{Marangotto:2020ead}. The leading uncertainties come from amplitude model choice and background contributions.

The largest contributions to the amplitude model, measured from fit fractions, come from the $\Kstarb(892)^0$, $\Deltares(1232)^{++}$ and $\Kstarbzz(1430)^0$ resonances. Among the $\Lz$ resonances, the largest contribution is from the $\Lz(2000)$ state, which is well described by the spin-$1/2^-$ Breit--Wigner lineshape determined in the \Lcpkpi analysis~\cite{LHCb-PAPER-2022-002}. Differences in resonance content and fit fractions of the \Xicpkpi baseline model with respect to \Lcpkpi~\cite{LHCb-PAPER-2022-002} show how the underlying dynamics of the two decays is different. The higher contribution of excited $\Lz$ and $K^*$ states in the \Xicpkpi case can be qualitatively explained by the presence of a valence \squark quark in the \Xicp baryon, which can be directly passed to the strange resonances. In the \Lcpkpi case, the strangeness of intermediate resonances has to be produced in the \Lc weak decay.

A large sensitivity of the \Xicpkpi decay to the polarization is measured, \mbox{$\sqrt{3}S = 0.691 \pm 0.005 \pm 0.030$}, which is also an observation of parity violation. This value is comparable with the sensitivity to \Lc polarization measured for the \Lcpkpi decay~\cite{LHCb-PAPER-2022-002}, $\sqrt{3}S \approx 0.662$. The large sensitivity makes the decay suitable for \Xicp polarization measurements in other systems. 
Many two-body decay-asymmetry parameters are significantly different from zero, indicating parity violation in these resonant decay contributions.

\begin{table}
\centering
\caption{Measured \Xicp polarization components and magnitude in $\tilde{B}$ and \textit{lab} frames.\label{tab:results_final_pol}}
\begin{tabular}{lcccc}
\toprule
Component & Central value (\%) & Stat. unc. & Model unc. & Syst. unc. \\
\midrule
$P_x$ ($\tilde{B}$) & \phantom{$-$}21.8\phantom{0} & 0.7\phantom{0} & 1.1\phantom{0} & 0.8\phantom{0} \\
$P_y$ ($\tilde{B}$) & $-0.73$ & 0.69 & 0.23 & 0.35 \\
$P_z$ ($\tilde{B}$) & $-$61.9\phantom{0} & 0.8\phantom{0} & 2.8\phantom{0} & 1.5\phantom{0} \\
$P$ ($\tilde{B}$) & \phantom{$-$}65.6\phantom{0} & 0.8\phantom{0} & 3.0\phantom{0} & 1.6\phantom{0} \\
\midrule
$P_x$ (\textit{lab}) & \phantom{$-$}56.9\phantom{0} & 0.7\phantom{0} & 2.6\phantom{0} & 1.0\phantom{0} \\
$P_y$ (\textit{lab}) & $-$0.41 & 0.68 & 0.19 & 0.23 \\
$P_z$ (\textit{lab}) & $-$26.2\phantom{0} & 0.7\phantom{0} & 1.4\phantom{0} & 2.0\phantom{0} \\
$P$ (\textit{lab}) & \phantom{$-$}62.6\phantom{0} & 0.7\phantom{0} & 3.0\phantom{0} & 1.2\phantom{0} \\
\bottomrule
\end{tabular}
\end{table}

\begin{table}
\centering
\caption{Baseline amplitude-model measured fit parameters of the $\Lz$ contributions with lower mass.\label{tab:results_final_Llow}}
\begin{tabular}{lccccccc}
\toprule
Parameter & Central value & Stat. unc. & Model unc. & Syst. unc. \\
\midrule
\Real$\mathcal{H}^{\Lz(1405)}_{1/2,0}$ & $-$0.74 & 0.22 & 0.69 & 0.17 \\
\Imag$\mathcal{H}^{\Lz(1405)}_{1/2,0}$ & \phantom{$-$}0.74 & 0.13 & 0.60 & 0.44 \\
\Real$\mathcal{H}^{\Lz(1405)}_{-1/2,0}$ & \phantom{$-$}2.5\phantom{0} & 0.1\phantom{0} & 1.0\phantom{0} & 0.4\phantom{0} \\
\Imag$\mathcal{H}^{\Lz(1405)}_{-1/2,0}$ & \phantom{$-$}1.1\phantom{0} & 0.2\phantom{0} & 1.5\phantom{0} & 0.6\phantom{0} \\
\midrule
\Real$\mathcal{H}^{\Lz(1520)}_{1/2,0}$ & $-$0.139 & 0.022 & 0.059 & 0.052 \\
\Imag$\mathcal{H}^{\Lz(1520)}_{1/2,0}$ & $-$0.182 & 0.025 & 0.025 & 0.030 \\
\Real$\mathcal{H}^{\Lz(1520)}_{-1/2,0}$ & $-$0.12\phantom{0} & 0.04\phantom{0} & 0.13\phantom{0} & 0.15\phantom{0} \\
\Imag$\mathcal{H}^{\Lz(1520)}_{-1/2,0}$ & \phantom{$-$}0.620 & 0.021 & 0.065 & 0.020 \\
\midrule
\Real$\mathcal{H}^{\Lz(1600)}_{1/2,0}$ & \phantom{$-$}1.20 & 0.12 & 0.52 & 0.25 \\
\Imag$\mathcal{H}^{\Lz(1600)}_{1/2,0}$ & \phantom{$-$}0.03 & 0.12 & 0.62 & 0.30 \\
\Real$\mathcal{H}^{\Lz(1600)}_{-1/2,0}$ & \phantom{$-$}0.51 & 0.13 & 0.73 & 0.64 \\
\Imag$\mathcal{H}^{\Lz(1600)}_{-1/2,0}$ & \phantom{$-$}1.16 & 0.12 & 0.90 & 0.17 \\
\midrule
\Real$\mathcal{H}^{\Lz(1670)}_{1/2,0}$ & $-$0.121 & 0.021 & 0.044 & 0.041 \\
\Imag$\mathcal{H}^{\Lz(1670)}_{1/2,0}$ & $-$0.020 & 0.022 & 0.063 & 0.017 \\
\Real$\mathcal{H}^{\Lz(1670)}_{-1/2,0}$ & $-$0.261 & 0.017 & 0.029 & 0.016 \\
\Imag$\mathcal{H}^{\Lz(1670)}_{-1/2,0}$ & $-$0.076 & 0.025 & 0.085 & 0.013 \\
\midrule
\Real$\mathcal{H}^{\Lz(1690)}_{1/2,0}$ & \phantom{$-$}0.15\phantom{0} & 0.06\phantom{0} & 0.21\phantom{0} & 0.14\phantom{0} \\
\Imag$\mathcal{H}^{\Lz(1690)}_{1/2,0}$ & $-$0.493 & 0.054 & 0.078 & 0.089 \\
\Real$\mathcal{H}^{\Lz(1690)}_{-1/2,0}$ & $-$0.85\phantom{0} & 0.05\phantom{0} & 0.24\phantom{0} & 0.04\phantom{0} \\
\Imag$\mathcal{H}^{\Lz(1690)}_{-1/2,0}$ & \phantom{$-$}0.52\phantom{0} & 0.06\phantom{0} & 0.19\phantom{0} & 0.11\phantom{0} \\
\midrule
\Real$\mathcal{H}^{\Lz(1710)}_{1/2,0}$ & \phantom{$-$}0.4\phantom{0} & 0.1\phantom{0} & 1.0\phantom{0} & 0.1\phantom{0} \\
\Imag$\mathcal{H}^{\Lz(1710)}_{1/2,0}$ & $-$0.26 & 0.15 & 0.30 & 0.10 \\
\Real$\mathcal{H}^{\Lz(1710)}_{-1/2,0}$ & \phantom{$-$}1.59 & 0.13 & 0.53 & 0.12 \\
\Imag$\mathcal{H}^{\Lz(1710)}_{-1/2,0}$ & \phantom{$-$}0.36 & 0.17 & 0.75 & 0.48 \\
\bottomrule
\end{tabular}
\end{table}

\begin{table}
\centering
\caption{Baseline amplitude-model measured fit parameters of the $\Lz$ contributions with higher mass.\label{tab:results_final_Lhigh}}
\begin{tabular}{lccccccc}
\toprule
Parameter & Central value & Stat. unc. & Model unc. & Syst. unc. \\
\midrule
\Real$\mathcal{H}^{\Lz(1800)}_{1/2,0}$ & \phantom{$-$}0.19 & 0.13 & 0.62 & 0.12 \\
\Imag$\mathcal{H}^{\Lz(1800)}_{1/2,0}$ & $-$0.73 & 0.12 & 0.46 & 0.20 \\
\Real$\mathcal{H}^{\Lz(1800)}_{-1/2,0}$ & $-$0.81 & 0.12 & 0.62 & 0.14 \\
\Imag$\mathcal{H}^{\Lz(1800)}_{-1/2,0}$ & $-$0.96 & 0.13 & 0.65 & 0.23 \\
\midrule
\Real$\mathcal{H}^{\Lz(1810)}_{1/2,0}$ & \phantom{$-$}0.05 & 0.10 & 0.36 & 0.07 \\
\Imag$\mathcal{H}^{\Lz(1810)}_{1/2,0}$ & \phantom{$-$}0.97 & 0.09 & 0.40 & 0.06 \\
\Real$\mathcal{H}^{\Lz(1810)}_{-1/2,0}$ & $-$0.08 & 0.08 & 0.21 & 0.08 \\
\Imag$\mathcal{H}^{\Lz(1810)}_{-1/2,0}$ & \phantom{$-$}0.11 & 0.08 & 0.15 & 0.09 \\
\midrule
\Real$\mathcal{H}^{\Lz(1820)}_{1/2,0}$ & \phantom{$-$}0.21 & 0.08 & 0.32 & 0.04 \\
\Imag$\mathcal{H}^{\Lz(1820)}_{1/2,0}$ & \phantom{$-$}1.05 & 0.07 & 0.18 & 0.08 \\
\Real$\mathcal{H}^{\Lz(1820)}_{-1/2,0}$ & $-$0.50 & 0.06 & 0.17 & 0.14 \\
\Imag$\mathcal{H}^{\Lz(1820)}_{-1/2,0}$ & \phantom{$-$}0.04 & 0.06 & 0.21 & 0.18 \\
\midrule
\Real$\mathcal{H}^{\Lz(1830)}_{1/2,0}$ & \phantom{$-$}0.50 & 0.07 & 0.56 & 0.08 \\
\Imag$\mathcal{H}^{\Lz(1830)}_{1/2,0}$ & $-$0.04 & 0.07 & 0.60 & 0.09 \\
\Real$\mathcal{H}^{\Lz(1830)}_{-1/2,0}$ & $-$0.21 & 0.07 & 0.27 & 0.10 \\
\Imag$\mathcal{H}^{\Lz(1830)}_{-1/2,0}$ & $-$0.30 & 0.07 & 0.26 & 0.10 \\
\midrule
\Real$\mathcal{H}^{\Lz(1890)}_{1/2,0}$ & \phantom{$-$}0.32 & 0.07 & 0.36 & 0.06 \\
\Imag$\mathcal{H}^{\Lz(1890)}_{1/2,0}$ & $-$0.19 & 0.07 & 0.19 & 0.07 \\
\Real$\mathcal{H}^{\Lz(1890)}_{-1/2,0}$ & \phantom{$-$}0.07 & 0.07 & 0.14 & 0.05 \\
\Imag$\mathcal{H}^{\Lz(1890)}_{-1/2,0}$ & $-$0.45 & 0.06 & 0.53 & 0.07 \\
\midrule
\Real$\mathcal{H}^{\Lz(2000)}_{1/2,0}$ & $-$1.45 & 0.14 & 0.36 & 0.17 \\
\Imag$\mathcal{H}^{\Lz(2000)}_{1/2,0}$ & $-$2.64 & 0.14 & 0.79 & 0.28 \\
\Real$\mathcal{H}^{\Lz(2000)}_{-1/2,0}$ & $-$0.71 & 0.12 & 0.29 & 0.07 \\
\Imag$\mathcal{H}^{\Lz(2000)}_{-1/2,0}$ & $-$1.52 & 0.10 & 0.50 & 0.08 \\
\bottomrule
\end{tabular}
\end{table}

\begin{table}
\centering
\caption{Baseline amplitude-model measured fit parameters of the $K^*$ contributions.\label{tab:results_final_K}}
\begin{tabular}{lccccccc}
\toprule
Parameter & Central value & Stat. unc. & Model unc. & Syst. unc. \\
\midrule 
\Real$\mathcal{H}^{\Kstarbzz(700)^0}_{1/2,0}$ & $-$3.95 & 0.23 & 0.86 & 0.37 \\
\Imag$\mathcal{H}^{\Kstarbzz(700)^0}_{1/2,0}$ & \phantom{$-$}3.0\phantom{0} & 0.3\phantom{0} & 1.5\phantom{0} & 0.6\phantom{0} \\
\Real$\mathcal{H}^{\Kstarbzz(700)^0}_{-1/2,0}$ & $-$2.5\phantom{0} & 0.2\phantom{0} & 1.2\phantom{0} & 0.5\phantom{0} \\
\Imag$\mathcal{H}^{\Kstarbzz(700)^0}_{-1/2,0}$ & $-$0.35 & 0.19 & 0.99 & 0.22 \\
$\gamma^{\Kstarbzz(700)^0} \left[\gev^{-2}\right]$ & $-$0.77 & 0.08 & 0.59 & 0.23 \\
\midrule
\Real$\mathcal{H}^{\Kstarb(892)^0}_{1/2,0}$ & \phantom{$-$}1\phantom{.000} & (fixed)\\
\Imag$\mathcal{H}^{\Kstarb(892)^0}_{1/2,0}$ & \phantom{$-$}0\phantom{.000} & (fixed)\\
\Real$\mathcal{H}^{\Kstarb(892)^0}_{1/2,-1}$ & \phantom{$-$}1.90\phantom{0} & 0.05\phantom{0} & 0.22\phantom{0} & 0.07\phantom{0} \\
\Imag$\mathcal{H}^{\Kstarb(892)^0}_{1/2,-1}$ & \phantom{$-$}0.52\phantom{0} & 0.08\phantom{0} & 0.22\phantom{0} & 0.13\phantom{0} \\
\Real$\mathcal{H}^{\Kstarb(892)^0}_{-1/2,1}$ & $-$1.037 & 0.044 & 0.038 & 0.043 \\
\Imag$\mathcal{H}^{\Kstarb(892)^0}_{-1/2,1}$ & \phantom{$-$}0.19\phantom{0} & 0.06\phantom{0} & 0.24\phantom{0} & 0.05\phantom{0} \\
\Real$\mathcal{H}^{\Kstarb(892)^0}_{-1/2,0}$ & $-$0.24\phantom{0} & 0.08\phantom{0} & 0.23\phantom{0} & 0.03\phantom{0} \\
\Imag$\mathcal{H}^{\Kstarb(892)^0}_{-1/2,0}$ & $-$1.20\phantom{0} & 0.05\phantom{0} & 0.17\phantom{0} & 0.02\phantom{0} \\
\midrule
\Real$\mathcal{H}^{\Kstarbzz(1430)^0}_{1/2,0}$ & \phantom{$-$}1.3\phantom{0} & 0.1\phantom{0} & 1.2\phantom{0} & 0.5\phantom{0} \\
\Imag$\mathcal{H}^{\Kstarbzz(1430)^0}_{1/2,0}$ & \phantom{$-$}1.0\phantom{0} & 0.1\phantom{0} & 1.4\phantom{0} & 0.3\phantom{0} \\
\Real$\mathcal{H}^{\Kstarbzz(1430)^0}_{-1/2,0}$ & $-$1.6\phantom{0} & 0.2\phantom{0} & 3.1\phantom{0} & 0.3\phantom{0} \\
\Imag$\mathcal{H}^{\Kstarbzz(1430)^0}_{-1/2,0}$ & \phantom{$-$}4.2\phantom{0} & 0.2\phantom{0} & 1.2\phantom{0} & 0.8\phantom{0} \\
$\gamma^{\Kstarbzz(1430)^0} \left[\gev^{-2}\right]$ & \phantom{$-$}0.05 & 0.02 & 0.27 & 0.07 \\
\midrule
\Real$\mathcal{H}^{\Kstarbtwo(1430)^0}_{1/2,0}$ & $-$1.3\phantom{0} & 0.1\phantom{0} & 1.1\phantom{0} & 0.3\phantom{0} \\
\Imag$\mathcal{H}^{\Kstarbtwo(1430)^0}_{1/2,0}$ & \phantom{$-$}0.31 & 0.11 & 0.59 & 0.29 \\
\Real$\mathcal{H}^{\Kstarbtwo(1430)^0}_{1/2,-1}$ & $-$0.54 & 0.17 & 0.17 & 0.23 \\
\Imag$\mathcal{H}^{\Kstarbtwo(1430)^0}_{1/2,-1}$ & \phantom{$-$}2.49 & 0.13 & 0.88 & 0.18 \\
\Real$\mathcal{H}^{\Kstarbtwo(1430)^0}_{-1/2,1}$ & \phantom{$-$}1.47 & 0.12 & 0.41 & 0.12 \\
\Imag$\mathcal{H}^{\Kstarbtwo(1430)^0}_{-1/2,1}$ & \phantom{$-$}0.85 & 0.15 & 0.65 & 0.16 \\
\Real$\mathcal{H}^{\Kstarbtwo(1430)^0}_{-1/2,0}$ & \phantom{$-$}0.95 & 0.11 & 0.29 & 0.13 \\
\Imag$\mathcal{H}^{\Kstarbtwo(1430)^0}_{-1/2,0}$ & $-$0.48 & 0.11 & 0.44 & 0.11 \\
\bottomrule
\end{tabular}
\end{table}

\begin{table}
\centering
\caption{Baseline amplitude-model measured fit parameters of the $\Deltares^{++}$ contributions.\label{tab:results_final_D}}
\begin{tabular}{lccccccc}
\toprule
Parameter & Central value & Stat. unc. & Model unc. & Syst. unc. \\
\midrule
\Real$\mathcal{H}^{\Deltares(1232)^{++}}_{1/2,0}$ & $-$1.50 & 0.08 & 0.14 & 0.06 \\
\Imag$\mathcal{H}^{\Deltares(1232)^{++}}_{1/2,0}$ & $-$0.07 & 0.10 & 0.25 & 0.14 \\
\Real$\mathcal{H}^{\Deltares(1232)^{++}}_{-1/2,0}$ & $-$3.32 & 0.16 & 0.32 & 0.05 \\
\Imag$\mathcal{H}^{\Deltares(1232)^{++}}_{-1/2,0}$ & \phantom{$-$}2.58 & 0.79 & 0.29 & 0.27 \\
\midrule
\Real$\mathcal{H}^{\Deltares(1600)^{++}}_{1/2,0}$ & \phantom{$-$}0.3\phantom{0} & 0.2\phantom{0} & 1.2\phantom{0} & 0.6\phantom{0} \\
\Imag$\mathcal{H}^{\Deltares(1600)^{++}}_{1/2,0}$ & $-$2.99 & 0.15 & 0.71 & 0.33 \\
\Real$\mathcal{H}^{\Deltares(1600)^{++}}_{-1/2,0}$ & \phantom{$-$}0.0\phantom{0} & 0.2\phantom{0} & 1.1\phantom{0} & 0.3\phantom{0} \\
\Imag$\mathcal{H}^{\Deltares(1600)^{++}}_{-1/2,0}$ & $-$2.08 & 0.13 & 0.30 & 0.35 \\
\midrule
\Real$\mathcal{H}^{\Deltares(1620)^{++}}_{1/2,0}$ & \phantom{$-$}0.00 & 0.08 & 0.33 & 0.10 \\
\Imag$\mathcal{H}^{\Deltares(1620)^{++}}_{1/2,0}$ & \phantom{$-$}1.20 & 0.05 & 0.40 & 0.05 \\
\Real$\mathcal{H}^{\Deltares(1620)^{++}}_{-1/2,0}$ & $-$0.67 & 0.06 & 0.28 & 0.21 \\
\Imag$\mathcal{H}^{\Deltares(1620)^{++}}_{-1/2,0}$ & $-$0.62 & 0.07 & 0.22 & 0.08 \\
\midrule
\Real$\mathcal{H}^{\Deltares(1700)^{++}}_{1/2,0}$ & \phantom{$-$}1.1\phantom{0} & 0.2\phantom{0} & 1.1\phantom{0} & 0.1\phantom{0} \\
\Imag$\mathcal{H}^{\Deltares(1700)^{++}}_{1/2,0}$ & $-$1.87 & 0.13 & 0.72 & 0.25 \\
\Real$\mathcal{H}^{\Deltares(1700)^{++}}_{-1/2,0}$ & \phantom{$-$}1.49 & 0.13 & 0.50 & 0.60 \\
\Imag$\mathcal{H}^{\Deltares(1700)^{++}}_{-1/2,0}$ & $-$1.09 & 0.13 & 0.50 & 0.13 \\
\bottomrule
\end{tabular}
\end{table}

\begin{table}
\centering
\caption{Fit fractions in \% of the resonant contributions included in the baseline amplitude model.\label{tab:fitfractions_final}}
\begin{tabular}{lccccccc}
\toprule
Resonance & Central value & Stat. unc. & Model unc. & Syst. unc. \\
\midrule 
$\Lz(1405)$ & \phantom{0}3.3\phantom{00} & 0.2\phantom{00} & 1.5\phantom{00} & 0.2\phantom{00} \\
$\Lz(1520)$ & \phantom{0}2.64\phantom{0} & 0.08\phantom{0} & 0.11\phantom{0} & 0.04\phantom{0} \\
$\Lz(1600)$ & \phantom{0}2.0\phantom{00} & 0.3\phantom{00} & 1.4\phantom{00} & 1.0\phantom{00} \\
$\Lz(1670)$ & \phantom{0}3.03\phantom{0} & 0.09\phantom{0} & 0.17\phantom{0} & 0.11\phantom{0} \\
$\Lz(1690)$ & \phantom{0}1.55\phantom{0} & 0.12\phantom{0} & 0.40\phantom{0} & 0.42\phantom{0} \\
$\Lz(1710)$ & \phantom{0}2.3\phantom{00} & 0.3\phantom{00} & 1.8\phantom{00} & 0.4\phantom{00} \\
$\Lz(1800)$ & \phantom{0}1.48\phantom{0} & 0.13\phantom{0} & 0.58\phantom{0} & 0.15\phantom{0} \\
$\Lz(1810)$ & \phantom{0}1.33\phantom{0} & 0.28\phantom{0} & 0.96\phantom{0} & 0.17\phantom{0} \\
$\Lz(1820)$ & \phantom{0}0.82\phantom{0} & 0.09\phantom{0} & 0.14\phantom{0} & 0.08\phantom{0} \\
$\Lz(1830)$ & \phantom{0}0.20\phantom{0} & 0.05\phantom{0} & 0.10\phantom{0} & 0.03\phantom{0} \\
$\Lz(1890)$ & \phantom{0}0.19\phantom{0} & 0.05\phantom{0} & 0.17\phantom{0} & 0.04\phantom{0} \\
$\Lz(2000)$ & \phantom{0}7.4\phantom{00} & 0.3\phantom{00} & 1.1\phantom{00} & 0.8\phantom{00} \\
\midrule
$\Kstarbzz(700)^0$ & \phantom{0}7.4\phantom{00} & 0.4\phantom{00} & 4.8\phantom{00} & 0.7\phantom{00} \\
$\Kstarb(892)^0$ & 28.61\phantom{0} & 0.28\phantom{0} & 0.82\phantom{0} & 0.80\phantom{0} \\
$\Kstarbzz(1430)^0$ & 15.6\phantom{00} & 0.7\phantom{00} & 7.1\phantom{00} & 1.9\phantom{00} \\
$\Kstarbtwo(1430)^0$ & \phantom{0}3.3\phantom{00} & 0.2\phantom{00} & 2.7\phantom{00} & 0.7\phantom{00} \\
\midrule
$\Deltares(1232)^{++}$ & 17.2\phantom{00} & 0.4\phantom{00} & 1.3\phantom{00} & 0.5\phantom{00} \\
$\Deltares(1600)^{++}$ & \phantom{0}4.31\phantom{0} & 0.27\phantom{0} & 0.96\phantom{0} & 0.91\phantom{0} \\
$\Deltares(1620)^{++}$ & \phantom{0}3.29\phantom{0} & 0.21\phantom{0} & 0.98\phantom{0} & 0.27\phantom{0} \\
$\Deltares(1700)^{++}$ & \phantom{0}2.01\phantom{0} & 0.17\phantom{0} & 0.44\phantom{0} & 0.15\phantom{0} \\
\bottomrule
\end{tabular}
\end{table}

\begin{table}
\centering
\caption{Sensitivity to polarization $\sqrt{3}S$ and decay-asymmetry $\alpha$ parameters of single resonant contributions.\label{tab:alphapars_final}}
\begin{tabular}{lccccccc}
\toprule
Resonance & $\alpha$ & Stat. unc. & Model unc. & Syst. unc. \\
\midrule 
Model $\sqrt{3}S$ & 0.691\phantom{0} & 0.005\phantom{0} & 0.029\phantom{0} & 0.009\phantom{0} \\
\midrule
$\Kstarb(892)^0$ $\sqrt{3}S$ & 0.613\phantom{0} & 0.016\phantom{0} & 0.062\phantom{0} & 0.014\phantom{0} \\
$\Kstarbtwo(1430)^0$ $\sqrt{3}S$ & 0.36\phantom{00} & 0.05\phantom{00} & 0.15\phantom{00} & 0.08\phantom{00} \\
\midrule
$\Lz(1405)$ & $-$0.75\phantom{0} & 0.09\phantom{0} & 0.28\phantom{0} & 0.08\phantom{0} \\
$\Lz(1520)$ & $-$0.77\phantom{0} & 0.04\phantom{0} & 0.11\phantom{0} & 0.06\phantom{0} \\
$\Lz(1600)$ & $-$0.06\phantom{0} & 0.11\phantom{0} & 0.39\phantom{0} & 0.08\phantom{0} \\
$\Lz(1670)$ & $-$0.66\phantom{0} & 0.06\phantom{0} & 0.13\phantom{0} & 0.11\phantom{0} \\
$\Lz(1690)$ & $-$0.58\phantom{0} & 0.07\phantom{0} & 0.07\phantom{0} & 0.12\phantom{0} \\
$\Lz(1710)$ & $-$0.86\phantom{0} & 0.09\phantom{0} & 0.35\phantom{0} & 0.06\phantom{0} \\
$\Lz(1800)$ & $-$0.47\phantom{0} & 0.30\phantom{0} & 0.95\phantom{0} & 0.71\phantom{0} \\
$\Lz(1810)$ & \phantom{$-$}0.96\phantom{0} & 0.06\phantom{0} & 0.42\phantom{0} & 0.08\phantom{0} \\
$\Lz(1820)$ & \phantom{$-$}0.64\phantom{0} & 0.07\phantom{0} & 0.19\phantom{0} & 0.21\phantom{0} \\
$\Lz(1830)$ & \phantom{$-$}0.30\phantom{0} & 0.19\phantom{0} & 0.91\phantom{0} & 0.41\phantom{0} \\
$\Lz(1890)$ & $-$0.19\phantom{0} & 0.19\phantom{0} & 0.51\phantom{0} & 0.21\phantom{0} \\
$\Lz(2000)$ & \phantom{$-$}0.53\phantom{0} & 0.04\phantom{0} & 0.13\phantom{0} & 0.05\phantom{0} \\
\midrule
$\Kstarbzz(700)^0$ & \phantom{$-$}0.60\phantom{0} & 0.04\phantom{0} & 0.10\phantom{0} & 0.06\phantom{0} \\
$\Kstarbzz(1430)^0$ & $-$0.758 & 0.028 & 0.081 & 0.054 \\
\midrule
$\Deltares(1232)^{++}$ & $-$0.774 & 0.020 & 0.066 & 0.019 \\
$\Deltares(1600)^{++}$ & \phantom{$-$}0.35\phantom{0} & 0.06\phantom{0} & 0.27\phantom{0} & 0.06\phantom{0} \\
$\Deltares(1620)^{++}$ & \phantom{$-$}0.26\phantom{0} & 0.06\phantom{0} & 0.31\phantom{0} & 0.22\phantom{0} \\
$\Deltares(1700)^{++}$ & \phantom{$-$}0.15\phantom{0} & 0.07\phantom{0} & 0.17\phantom{0} & 0.24\phantom{0} \\
\bottomrule
\end{tabular}
\end{table}

\section{Summary}
\label{sec:summary}

An amplitude analysis of  \Xicpkpi decays with a measurement of the \Xicp polarization vector in semileptonic beauty-hadron decays is presented. Candidates are selected from \proton\proton collisions recorded with the \lhcb detector at center-of-mass energies of $7$, $8$ and $13\tev$, corresponding to an integrated luminosity of $9 \invfb$. All the parameters of the amplitude model and the baryon polarization have been measured; fit fractions and decay-asymmetry parameters for each two-body resonant contribution are also reported together with the effective three-body decay-asymmetry parameter of the \Xicpkpi decay.
The most important resonances contributing to the \Xicpkpi decay are the $\Kstarb(892)^0$, $\Deltares(1232)^{++}$, and $\Kstarbzz(1430)^0$ states. Among the $\Lz$ resonances the largest contribution is from the $\Lz(2000)$ state. A large \Xicp polarization of order $63-66\%$ is found, measured with absolute uncertainties of order $1\%$. The normal polarization, sensitive to time-reversal violation effects and final-state interactions, is compatible with zero.
A large sensitivity to the polarization is measured, making the amplitude model suitable for \Xicp polarization measurements in other systems. The amplitude model obtained provides a complete description of the \Xicpkpi decay, with applications ranging from new physics searches to low-energy QCD. Such applications include an increased sensitivity to angular analyses of semileptonic baryon decays, and measurements of the \Xicp polarization and electromagnetic dipole moments via spin precession.

%% file: acknowledgements.tex
\section*{Acknowledgements}
%
%
\noindent We express our gratitude to our colleagues in the CERN
accelerator departments for the excellent performance of the LHC. We
thank the technical and administrative staff at the LHCb
institutes.
We acknowledge support from CERN and from the national agencies:
ARC (Australia);
CAPES, CNPq, FAPERJ and FINEP (Brazil); 
MOST and NSFC (China); 
CNRS/IN2P3 (France); 
BMBF, DFG and MPG (Germany); 
INFN (Italy); 
NWO (Netherlands); 
MNiSW and NCN (Poland); 
MCID/IFA (Romania); 
MICIU and AEI (Spain);
SNSF and SER (Switzerland); 
NASU (Ukraine); 
STFC (United Kingdom); 
DOE NP and NSF (USA).
We acknowledge the computing resources that are provided by ARDC (Australia), 
CBPF (Brazil),
CERN, 
IHEP and LZU (China),
IN2P3 (France), 
KIT and DESY (Germany), 
INFN (Italy), 
SURF (Netherlands),
Polish WLCG (Poland),
IFIN-HH (Romania), 
PIC (Spain), CSCS (Switzerland), 
and GridPP (United Kingdom).
We are indebted to the communities behind the multiple open-source
software packages on which we depend.
Individual groups or members have received support from
Key Research Program of Frontier Sciences of CAS, CAS PIFI, CAS CCEPP, 
Fundamental Research Funds for the Central Universities,  and Sci.\ \& Tech.\ Program of Guangzhou (China);
Minciencias (Colombia);
EPLANET, Marie Sk\l{}odowska-Curie Actions, ERC and NextGenerationEU (European Union);
A*MIDEX, ANR, IPhU and Labex P2IO, and R\'{e}gion Auvergne-Rh\^{o}ne-Alpes (France);
Alexander-von-Humboldt Foundation (Germany);
ICSC (Italy); 
Severo Ochoa and Mar\'ia de Maeztu Units of Excellence, GVA, XuntaGal, GENCAT, InTalent-Inditex and Prog.~Atracci\'on Talento CM (Spain);
SRC (Sweden);
the Leverhulme Trust, the Royal Society and UKRI (United Kingdom).

%% file: appendix.tex
\clearpage

\appendix

\section{Summary of systematic uncertainty contributions}
\label{sec:syst_summary}

Systematic uncertainties separated for each contribution are reported in
Tables~\ref{tab:syst_fit_pars_Llow}-\ref{tab:syst_fit_pars_D} for fit parameters, Table~\ref{tab:syst_fitfraction} for fit fractions, and Table~\ref{tab:syst_alphapars} for decay asymmetries.

\begin{table}[b]
\centering
\caption{Systematic uncertainty contributions on fit parameters describing the $\Lz$ contributions with lower mass. Total* includes all contributions except for the choice of amplitude model. \label{tab:syst_fit_pars_Llow}}
\begin{tabular}{lcccccc}
\toprule
Parameter & Model & Total* & Background & Kinematics & PID & Fit Bias \\
\midrule 
\Real$\mathcal{H}^{\Lz(1405)}_{1/2,0}$ & 0.69 & 0.17 & 0.17 & $<$0.01 & \phantom{$<$}0.02 & $<$0.01 \\
\Imag$\mathcal{H}^{\Lz(1405)}_{1/2,0}$ & 0.60 & 0.44 & 0.41 & \phantom{$<$}0.15 & \phantom{$<$}0.01 & \phantom{$<$}0.01 \\
\Real$\mathcal{H}^{\Lz(1405)}_{-1/2,0}$ & 1.0\phantom{0} & 0.4\phantom{0} & 0.4\phantom{0} & $<$0.1\phantom{0} & $<$0.1\phantom{0} & $<$0.1\phantom{0} \\
\Imag$\mathcal{H}^{\Lz(1405)}_{-1/2,0}$ & 1.5\phantom{0} & 0.6\phantom{0} & 0.6\phantom{0} & $<$0.1\phantom{0} & $<$0.1\phantom{0} & $<$0.1\phantom{0} \\
\midrule
\Real$\mathcal{H}^{\Lz(1520)}_{1/2,0}$ & 0.059 & 0.052 & 0.051 & \phantom{$<$}0.003 & 0.009 & \phantom{$<$}0.002 \\
\Imag$\mathcal{H}^{\Lz(1520)}_{1/2,0}$ & 0.025 & 0.030 & 0.022 & \phantom{$<$}0.021 & 0.004 & \phantom{$<$}0.001 \\
\Real$\mathcal{H}^{\Lz(1520)}_{-1/2,0}$ & 0.13\phantom{0} & 0.15\phantom{0} & 0.15\phantom{0} & $<$0.01\phantom{0} & 0.01\phantom{0} & $<$0.01\phantom{0} \\
\Imag$\mathcal{H}^{\Lz(1520)}_{-1/2,0}$ & 0.065 & 0.020 & 0.019 & \phantom{$<$}0.007 & 0.002 & \phantom{$<$}0.001 \\
\midrule
\Real$\mathcal{H}^{\Lz(1600)}_{1/2,0}$ & 0.52 & 0.25 & 0.24 & 0.05 & 0.02 & \phantom{$<$}0.01 \\
\Imag$\mathcal{H}^{\Lz(1600)}_{1/2,0}$ & 0.62 & 0.30 & 0.29 & 0.01 & 0.06 & \phantom{$<$}0.01 \\
\Real$\mathcal{H}^{\Lz(1600)}_{-1/2,0}$ & 0.73 & 0.64 & 0.64 & 0.04 & 0.03 & $<$0.01 \\
\Imag$\mathcal{H}^{\Lz(1600)}_{-1/2,0}$ & 0.90 & 0.17 & 0.17 & 0.03 & 0.02 & \phantom{$<$}0.01 \\
\midrule
\Real$\mathcal{H}^{\Lz(1670)}_{1/2,0}$ & 0.044 & 0.041 & 0.039 & \phantom{$<$}0.010 & 0.007 & \phantom{$<$}0.002 \\
\Imag$\mathcal{H}^{\Lz(1670)}_{1/2,0}$ & 0.063 & 0.017 & 0.017 & $<$0.001 & 0.003 & $<$0.001 \\
\Real$\mathcal{H}^{\Lz(1670)}_{-1/2,0}$ & 0.029 & 0.016 & 0.012 & \phantom{$<$}0.007 & 0.008 & $<$0.001 \\
\Imag$\mathcal{H}^{\Lz(1670)}_{-1/2,0}$ & 0.085 & 0.013 & 0.009 & \phantom{$<$}0.008 & 0.006 & \phantom{$<$}0.001 \\
\midrule
\Real$\mathcal{H}^{\Lz(1690)}_{1/2,0}$ & 0.21\phantom{0} & 0.14\phantom{0} & 0.13\phantom{0} & \phantom{$<$}0.03\phantom{0} & 0.01\phantom{0} & $<$0.01\phantom{0} \\
\Imag$\mathcal{H}^{\Lz(1690)}_{1/2,0}$ & 0.079 & 0.089 & 0.084 & \phantom{$<$}0.030 & 0.004 & \phantom{$<$}0.001 \\
\Real$\mathcal{H}^{\Lz(1690)}_{-1/2,0}$ & 0.24\phantom{0} & 0.04\phantom{0} & 0.04\phantom{0} & $<$0.01\phantom{0} & 0.01\phantom{0} & $<$0.01\phantom{0} \\
\Imag$\mathcal{H}^{\Lz(1690)}_{-1/2,0}$ & 0.19\phantom{0} & 0.11\phantom{0} & 0.11\phantom{0} & \phantom{$<$}0.01\phantom{0} & 0.01\phantom{0} & $<$0.01\phantom{0} \\
\midrule
\Real$\mathcal{H}^{\Lz(1710)}_{1/2,0}$ & 1.0\phantom{0} & 0.1\phantom{0} & 0.1\phantom{0} & 0.1\phantom{0} & 0.1\phantom{0} & $<$0.1\phantom{0} \\
\Imag$\mathcal{H}^{\Lz(1710)}_{1/2,0}$ & 0.30 & 0.10 & 0.09 & 0.03 & 0.02 & $<$0.01 \\
\Real$\mathcal{H}^{\Lz(1710)}_{-1/2,0}$ & 0.53 & 0.12 & 0.12 & 0.03 & 0.02 & \phantom{$<$}0.01 \\
\Imag$\mathcal{H}^{\Lz(1710)}_{-1/2,0}$ & 0.75 & 0.48 & 0.48 & 0.02 & 0.04 & $<$0.01 \\
\bottomrule
\end{tabular}
\end{table}
\begin{table}
\centering
\caption{Systematic uncertainty contributions on fit parameters describing the $\Lz$ contributions with higher mass. Total* includes all contributions except for the choice of amplitude model.\label{tab:syst_fit_pars_Lhigh}}
\begin{tabular}{lcccccc}
\toprule
Parameter & Model & Total* & Background & Kinematics & PID & Fit Bias \\
\midrule
\Real$\mathcal{H}^{\Lz(1800)}_{1/2,0}$ & 0.62 & 0.12 & 0.11 & 0.04 & 0.01 & $<$0.01 \\
\Imag$\mathcal{H}^{\Lz(1800)}_{1/2,0}$ & 0.46 & 0.20 & 0.19 & 0.07 & 0.01 & \phantom{$<$}0.01 \\
\Real$\mathcal{H}^{\Lz(1800)}_{-1/2,0}$ & 0.62 & 0.14 & 0.14 & 0.02 & 0.01 & \phantom{$<$}0.01 \\
\Imag$\mathcal{H}^{\Lz(1800)}_{-1/2,0}$ & 0.65 & 0.23 & 0.23 & 0.01 & 0.01 & $<$0.01 \\
\midrule
\Real$\mathcal{H}^{\Lz(1810)}_{1/2,0}$ & 0.36 & 0.07 & 0.06 & \phantom{$<$}0.03 & 0.01 & $<$0.01 \\
\Imag$\mathcal{H}^{\Lz(1810)}_{1/2,0}$ & 0.40 & 0.06 & 0.05 & $<$0.01 & 0.02 & $<$0.01 \\
\Real$\mathcal{H}^{\Lz(1810)}_{-1/2,0}$ & 0.21 & 0.08 & 0.07 & \phantom{$<$}0.04 & 0.01 & $<$0.01 \\
\Imag$\mathcal{H}^{\Lz(1810)}_{-1/2,0}$ & 0.20 & 0.09 & 0.08 & \phantom{$<$}0.02 & 0.02 & $<$0.01 \\
\midrule
\Real$\mathcal{H}^{\Lz(1820)}_{1/2,0}$ & 0.32 & 0.04 & 0.04 & \phantom{$<$}0.01 & \phantom{$<$}0.01 & $<$0.01 \\
\Imag$\mathcal{H}^{\Lz(1820)}_{1/2,0}$ & 0.18 & 0.08 & 0.07 & $<$0.01 & \phantom{$<$}0.02 & $<$0.00 \\
\Real$\mathcal{H}^{\Lz(1820)}_{-1/2,0}$ & 0.17 & 0.14 & 0.13 & \phantom{$<$}0.02 & $<$0.01 & $<$0.01 \\
\Imag$\mathcal{H}^{\Lz(1820)}_{-1/2,0}$ & 0.21 & 0.18 & 0.18 & \phantom{$<$}0.01 & \phantom{$<$}0.01 & $<$0.01 \\
\midrule
\Real$\mathcal{H}^{\Lz(1830)}_{1/2,0}$ & 0.56 & 0.08 & 0.07 & \phantom{$<$}0.01 & 0.01 & $<$0.01 \\
\Imag$\mathcal{H}^{\Lz(1830)}_{1/2,0}$ & 0.59 & 0.09 & 0.09 & $<$0.01 & 0.01 & $<$0.01 \\
\Real$\mathcal{H}^{\Lz(1830)}_{-1/2,0}$ & 0.27 & 0.10 & 0.09 & \phantom{$<$}0.02 & 0.04 & $<$0.01 \\
\Imag$\mathcal{H}^{\Lz(1830)}_{-1/2,0}$ & 0.26 & 0.10 & 0.09 & \phantom{$<$}0.03 & 0.03 & $<$0.01 \\
\midrule
\Real$\mathcal{H}^{\Lz(1890)}_{1/2,0}$ & 0.36 & 0.06 & 0.05 & \phantom{$<$}0.02 & 0.01 & $<$0.01 \\
\Imag$\mathcal{H}^{\Lz(1890)}_{1/2,0}$ & 0.19 & 0.08 & 0.07 & \phantom{$<$}0.02 & 0.01 & \phantom{$<$}0.01 \\
\Real$\mathcal{H}^{\Lz(1890)}_{-1/2,0}$ & 0.14 & 0.05 & 0.05 & $<$0.01 & 0.02 & $<$0.01 \\
\Imag$\mathcal{H}^{\Lz(1890)}_{-1/2,0}$ & 0.53 & 0.07 & 0.07 & $<$0.01 & 0.01 & $<$0.01 \\
\midrule
\Real$\mathcal{H}^{\Lz(2000)}_{1/2,0}$ & 0.36 & 0.17 & 0.11 & 0.12 & 0.02 & $<$0.01 \\
\Imag$\mathcal{H}^{\Lz(2000)}_{1/2,0}$ & 0.79 & 0.28 & 0.27 & 0.02 & 0.03 & $<$0.01 \\
\Real$\mathcal{H}^{\Lz(2000)}_{-1/2,0}$ & 0.29 & 0.07 & 0.07 & 0.02 & 0.01 & \phantom{$<$}0.01 \\
\Imag$\mathcal{H}^{\Lz(2000)}_{-1/2,0}$ & 0.50 & 0.08 & 0.07 & 0.02 & 0.02 & $<$0.01 \\
\bottomrule
\end{tabular}
\end{table}
\begin{table}
\centering
\caption{Systematic uncertainty contributions on fit parameters describing $K^*$ contributions. Total* includes all contributions except for the choice of amplitude model.\label{tab:syst_fit_pars_K}}
\begin{tabular}{lcccccc}
\toprule
Parameter & Model & Total* & Background & Kinematics & PID & Fit Bias \\
\midrule
\Real$\mathcal{H}^{\Kstarbzz(700)^0}_{1/2,0}$ & 0.86 & 0.37 & 0.36 & \phantom{$<$}0.08 & 0.07 & \phantom{$<$}0.02 \\
\Imag$\mathcal{H}^{\Kstarbzz(700)^0}_{1/2,0}$ & 1.5\phantom{0} & 0.6\phantom{0} & 0.5\phantom{0} & \phantom{$<$}0.2\phantom{0} & 0.1\phantom{0} & $<$0.1\phantom{0} \\
\Real$\mathcal{H}^{\Kstarbzz(700)^0}_{-1/2,0}$ & 1.2\phantom{0} & 0.5\phantom{0} & 0.5\phantom{0} & $<$0.1\phantom{0} & 0.1\phantom{0} & \phantom{$<$}0.1\phantom{0} \\
\Imag$\mathcal{H}^{\Kstarbzz(700)^0}_{-1/2,0}$ & 0.99 & 0.22 & 0.21 & \phantom{$<$}0.02 & 0.04 & \phantom{$<$}0.01 \\
$\gamma^{\Kstarbzz(700)^0} \left[\gev^{-2}\right]$ & 0.59 & 0.23 & 0.23 & \phantom{$<$}0.01 & 0.03 & $<$0.01 \\
\midrule
\Real$\mathcal{H}^{\Kstarb(892)^0}_{1/2,0}$ & \multicolumn{6}{c}{0 (fixed)}\\
\Imag$\mathcal{H}^{\Kstarb(892)^0}_{1/2,0}$ & \multicolumn{6}{c}{0 (fixed)}\\
\Real$\mathcal{H}^{\Kstarb(892)^0}_{1/2,-1}$ & 0.22\phantom{0} & 0.08\phantom{0} & 0.06\phantom{0} & 0.03\phantom{0} & 0.02\phantom{0} & \phantom{$<$}0.01\phantom{0} \\
\Imag$\mathcal{H}^{\Kstarb(892)^0}_{1/2,-1}$ & 0.22\phantom{0} & 0.13\phantom{0} & 0.12\phantom{0} & 0.03\phantom{0} & 0.01\phantom{0} & \phantom{$<$}0.01\phantom{0} \\
\Real$\mathcal{H}^{\Kstarb(892)^0}_{-1/2,1}$ & 0.038 & 0.043 & 0.043 & 0.001 & 0.003 & \phantom{$<$}0.003 \\
\Imag$\mathcal{H}^{\Kstarb(892)^0}_{-1/2,1}$ & 0.24\phantom{0} & 0.05\phantom{0} & 0.04\phantom{0} & 0.03\phantom{0} & 0.01\phantom{0} & $<$0.01\phantom{0} \\
\Real$\mathcal{H}^{\Kstarb(892)^0}_{-1/2,0}$ & 0.23\phantom{0} & 0.03\phantom{0} & 0.02\phantom{0} & 0.02\phantom{0} & 0.01\phantom{0} & $<$0.01\phantom{0} \\
\Imag$\mathcal{H}^{\Kstarb(892)^0}_{-1/2,0}$ & 0.17\phantom{0} & 0.02\phantom{0} & 0.02\phantom{0} & 0.01\phantom{0} & 0.01\phantom{0} & \phantom{$<$}0.01\phantom{0} \\
\midrule
\Real$\mathcal{H}^{\Kstarbzz(1430)^0}_{1/2,0}$ & 1.2 & 0.5 & 0.5 & \phantom{$<$}0.1 & \phantom{$<$}0.1 & $<$0.1 \\
\Imag$\mathcal{H}^{\Kstarbzz(1430)^0}_{1/2,0}$ & 1.4 & 0.3 & 0.3 & $<$0.1 & $<$0.1 & $<$0.1 \\
\Real$\mathcal{H}^{\Kstarbzz(1430)^0}_{-1/2,0}$ & 3.1 & 0.3 & 0.3 & \phantom{$<$}0.1 & $<$0.1 & $<$0.1 \\
\Imag$\mathcal{H}^{\Kstarbzz(1430)^0}_{-1/2,0}$ & 1.2 & 0.8 & 0.8 & \phantom{$<$}0.1 & \phantom{$<$}0.1 & \phantom{$<$}0.1 \\
$\gamma^{\Kstarbzz(1430)^0} \left[\gev^{-2}\right]$ & 0.27 & 0.07 & 0.07 & $<$0.01 & $<$0.01 & $<$0.01 \\
\midrule
\Real$\mathcal{H}^{\Kstarbtwo(1430)^0}_{1/2,0}$ & 1.1\phantom{0} & 0.3\phantom{0} & 0.3\phantom{0} & $<$0.1\phantom{0} & $<$0.1\phantom{0} & $<$0.1\phantom{0} \\
\Imag$\mathcal{H}^{\Kstarbtwo(1430)^0}_{1/2,0}$ & 0.59 & 0.29 & 0.29 & \phantom{$<$}0.04 & \phantom{$<$}0.01 & \phantom{$<$}0.01 \\
\Real$\mathcal{H}^{\Kstarbtwo(1430)^0}_{1/2,-1}$ & 0.17 & 0.23 & 0.20 & \phantom{$<$}0.11 & \phantom{$<$}0.03 & $<$0.01 \\
\Imag$\mathcal{H}^{\Kstarbtwo(1430)^0}_{1/2,-1}$ & 0.88 & 0.18 & 0.16 & \phantom{$<$}0.07 & \phantom{$<$}0.01 & \phantom{$<$}0.01 \\
\Real$\mathcal{H}^{\Kstarbtwo(1430)^0}_{-1/2,1}$ & 0.41 & 0.12 & 0.08 & \phantom{$<$}0.07 & \phantom{$<$}0.05 & $<$0.01 \\
\Imag$\mathcal{H}^{\Kstarbtwo(1430)^0}_{-1/2,1}$ & 0.65 & 0.16 & 0.15 & \phantom{$<$}0.03 & \phantom{$<$}0.03 & $<$0.01 \\
\Real$\mathcal{H}^{\Kstarbtwo(1430)^0}_{-1/2,0}$ & 0.29 & 0.13 & 0.12 & \phantom{$<$}0.06 & \phantom{$<$}0.03 & $<$0.01 \\
\Imag$\mathcal{H}^{\Kstarbtwo(1430)^0}_{-1/2,0}$ & 0.44 & 0.11 & 0.10 & \phantom{$<$}0.03 & \phantom{$<$}0.03 & $<$0.01 \\
\bottomrule
\end{tabular}
\end{table}
\begin{table}
\centering
\caption{Systematic uncertainty contributions on fit parameters describing $\Deltares^{++}$ contributions. Total* includes all contributions except for the choice of amplitude model.\label{tab:syst_fit_pars_D}}
\begin{tabular}{lcccccc}
\toprule
Parameter & Model & Total* & Background & Kinematics & PID & Fit Bias \\
\midrule 
\Real$\mathcal{H}^{\Deltares(1232)^{++}}_{1/2,0}$ & 0.14 & 0.06 & 0.05 & $<$0.01 & 0.02 & $<$0.01 \\
\Imag$\mathcal{H}^{\Deltares(1232)^{++}}_{1/2,0}$ & 0.25 & 0.14 & 0.13 & \phantom{$<$}0.05 & 0.02 & $<$0.01 \\
\Real$\mathcal{H}^{\Deltares(1232)^{++}}_{-1/2,0}$ & 0.32 & 0.05 & 0.05 & \phantom{$<$}0.01 & 0.02 & \phantom{$<$}0.01 \\
\Imag$\mathcal{H}^{\Deltares(1232)^{++}}_{-1/2,0}$ & 0.79 & 0.27 & 0.26 & \phantom{$<$}0.07 & 0.06 & \phantom{$<$}0.01 \\
\midrule
\Real$\mathcal{H}^{\Deltares(1600)^{++}}_{1/2,0}$ & 1.2\phantom{0} & 0.6\phantom{0} & 0.6\phantom{0} & $<$0.1\phantom{0} & $<$0.1\phantom{0} & $<$0.1\phantom{0} \\
\Imag$\mathcal{H}^{\Deltares(1600)^{++}}_{1/2,0}$ & 0.71 & 0.33 & 0.33 & \phantom{$<$}0.03 & \phantom{$<$}0.05 & $<$0.01 \\
\Real$\mathcal{H}^{\Deltares(1600)^{++}}_{-1/2,0}$ & 1.1\phantom{0} & 0.3\phantom{0} & 0.2\phantom{0} & $<$0.1\phantom{0} & $<$0.1\phantom{0} & $<$0.1\phantom{0} \\
\Imag$\mathcal{H}^{\Deltares(1600)^{++}}_{-1/2,0}$ & 0.30 & 0.35 & 0.34 & \phantom{$<$}0.05 & \phantom{$<$}0.03 & $<$0.01 \\
\midrule
\Real$\mathcal{H}^{\Deltares(1620)^{++}}_{1/2,0}$ & 0.33 & 0.10 & 0.10 & $<$0.01 & \phantom{$<$}0.02 & $<$0.01 \\
\Imag$\mathcal{H}^{\Deltares(1620)^{++}}_{1/2,0}$ & 0.40 & 0.05 & 0.05 & \phantom{$<$}0.01 & \phantom{$<$}0.01 & $<$0.01 \\
\Real$\mathcal{H}^{\Deltares(1620)^{++}}_{-1/2,0}$ & 0.28 & 0.21 & 0.21 & \phantom{$<$}0.04 & \phantom{$<$}0.01 & $<$0.01 \\
\Imag$\mathcal{H}^{\Deltares(1620)^{++}}_{-1/2,0}$ & 0.22 & 0.08 & 0.08 & \phantom{$<$}0.01 & $<$0.01 & $<$0.01 \\
\midrule
\Real$\mathcal{H}^{\Deltares(1700)^{++}}_{1/2,0}$ & 1.1\phantom{0} & 0.1\phantom{0} & 0.1\phantom{0} & $<$0.1\phantom{0} & 0.1\phantom{0} & $<$0.1\phantom{0} \\
\Imag$\mathcal{H}^{\Deltares(1700)^{++}}_{1/2,0}$ & 0.72 & 0.25 & 0.23 & \phantom{$<$}0.03 & 0.09 & $<$0.01 \\
\Real$\mathcal{H}^{\Deltares(1700)^{++}}_{-1/2,0}$ & 0.50 & 0.60 & 0.59 & \phantom{$<$}0.04 & 0.08 & \phantom{$<$}0.01 \\
\Imag$\mathcal{H}^{\Deltares(1700)^{++}}_{-1/2,0}$ & 0.50 & 0.13 & 0.10 & \phantom{$<$}0.06 & 0.04 & $<$0.01 \\
\bottomrule
\end{tabular}
\end{table}
\begin{table}
\centering
\caption{Systematic uncertainty contributions on fit fractions. Total* includes all contributions except for the choice of the amplitude model.\label{tab:syst_fitfraction}}
\begin{tabular}{lcccccc}
\toprule
Resonance & Model & Total* & Background & Kinematics & PID & Fit Bias \\
\midrule 
$\Lz(1405)$ & 0.015\phantom{00} & 0.002\phantom{00} & 0.002\phantom{00} & $<$0.001\phantom{00} & $<$0.001\phantom{00} & $<$0.001\phantom{00} \\
$\Lz(1520)$ & 0.0011\phantom{0} & 0.0004\phantom{0} & 0.0004\phantom{0} & \phantom{$<$}0.0001\phantom{0} & $<$0.0001\phantom{0} & \phantom{$<$}0.0001\phantom{0} \\
$\Lz(1600)$ & 0.014\phantom{00} & 0.010\phantom{00} & 0.010\phantom{00} & \phantom{$<$}0.001\phantom{00} & \phantom{$<$}0.001\phantom{00} & $<$0.001\phantom{00} \\
$\Lz(1670)$ & 0.0017\phantom{0} & 0.0011\phantom{0} & 0.0011\phantom{0} & $<$0.0001\phantom{0} & \phantom{$<$}0.0001\phantom{0} & \phantom{$<$}0.0002\phantom{0} \\
$\Lz(1690)$ & 0.0040\phantom{0} & 0.0042\phantom{0} & 0.0042\phantom{0} & $<$0.0001\phantom{0} & $<$0.0001\phantom{0} & $<$0.0001\phantom{0} \\
$\Lz(1710)$ & 0.018\phantom{00} & 0.004\phantom{00} & 0.004\phantom{00} & $<$0.001\phantom{00} & \phantom{$<$}0.001\phantom{00} & \phantom{$<$}0.001\phantom{00} \\
$\Lz(1800)$ & 0.0058\phantom{0} & 0.0015\phantom{0} & 0.0014\phantom{0} & \phantom{$<$}0.0003\phantom{0} & \phantom{$<$}0.0001\phantom{0} & \phantom{$<$}0.0004\phantom{0} \\
$\Lz(1810)$ & 0.0096\phantom{0} & 0.0017\phantom{0} & 0.0012\phantom{0} & \phantom{$<$}0.0004\phantom{0} & \phantom{$<$}0.0011\phantom{0} & \phantom{$<$}0.0002\phantom{0} \\
$\Lz(1820)$ & 0.0014\phantom{0} & 0.0008\phantom{0} & 0.0008\phantom{0} & \phantom{$<$}0.0001\phantom{0} & \phantom{$<$}0.0002\phantom{0} & $<$0.0001\phantom{0} \\
$\Lz(1830)$ & 0.0010\phantom{0} & 0.0003\phantom{0} & 0.0003\phantom{0} & \phantom{$<$}0.0001\phantom{0} & \phantom{$<$}0.0001\phantom{0} & \phantom{$<$}0.0001\phantom{0} \\
$\Lz(1890)$ & 0.0017\phantom{0} & 0.0004\phantom{0} & 0.0004\phantom{0} & \phantom{$<$}0.0001\phantom{0} & \phantom{$<$}0.0001\phantom{0} & \phantom{$<$}0.0001\phantom{0} \\
$\Lz(2000)$ & 0.011\phantom{00} & 0.008\phantom{00} & 0.008\phantom{00} & \phantom{$<$}0.001\phantom{00} & \phantom{$<$}0.001\phantom{00} & $<$0.001\phantom{00} \\
\midrule
$\Kstarbzz(700)^0$ & 0.048\phantom{00} & 0.007\phantom{00} & 0.006\phantom{00} & $<$0.001\phantom{00} & \phantom{$<$}0.002\phantom{00} & $<$0.001\phantom{00} \\
$\Kstarb(892)^0$ & 0.0082\phantom{0} & 0.0080\phantom{0} & 0.0076\phantom{0} & \phantom{$<$}0.0024\phantom{0} & \phantom{$<$}0.0005\phantom{0} & \phantom{$<$}0.0005\phantom{0} \\
$\Kstarbzz(1430)^0$ & 0.071\phantom{00} & 0.019\phantom{00} & 0.018\phantom{00} & \phantom{$<$}0.003\phantom{00} & \phantom{$<$}0.004\phantom{00} & $<$0.001\phantom{00} \\
$\Kstarbtwo(1430)^0$ & 0.027\phantom{00} & 0.007\phantom{00} & 0.006\phantom{00} & \phantom{$<$}0.002\phantom{00} & \phantom{$<$}0.001\phantom{00} & $<$0.001\phantom{00} \\
\midrule
$\Deltares(1232)^{++}$ & 0.013\phantom{00} & 0.005\phantom{00} & 0.004\phantom{00} & \phantom{$<$}0.002\phantom{00} & \phantom{$<$}0.001\phantom{00} & $<$0.001\phantom{00} \\
$\Deltares(1600)^{++}$ & 0.0096\phantom{0} & 0.0091\phantom{0} & 0.0091\phantom{0} & \phantom{$<$}0.0001\phantom{0} & \phantom{$<$}0.0006\phantom{0} & \phantom{$<$}0.0002\phantom{0} \\
$\Deltares(1620)^{++}$ & 0.0098\phantom{0} & 0.0027\phantom{0} & 0.0027\phantom{0} & \phantom{$<$}0.0003\phantom{0} & \phantom{$<$}0.0003\phantom{0} & $<$0.0001\phantom{0} \\
$\Deltares(1700)^{++}$ & 0.0044\phantom{0} & 0.0015\phantom{0} & 0.0012\phantom{0} & \phantom{$<$}0.0003\phantom{0} & \phantom{$<$}0.0008\phantom{0} & \phantom{$<$}0.0003\phantom{0} \\
\bottomrule
\end{tabular}
\end{table}
\begin{table}
\centering
\caption{Systematic uncertainties on $\sqrt{3}S$ and decay-asymmetry parameters. Total* includes all contributions except for the choice of amplitude model.\label{tab:syst_alphapars}}
\begin{tabular}{lcccccc}
\toprule
$\alpha$ & Model & Total* & Background & Kinematics & PID & Fit Bias \\
\midrule 
Model $\sqrt{3}S$ & 0.029\phantom{0} & 0.009\phantom{0} & 0.008\phantom{0} & \phantom{$<$}0.001\phantom{0} & 0.003\phantom{0} & $<$0.001\phantom{0} \\
\midrule
$\Kstarb(892)^0$ $\sqrt{3}S$ & 0.062\phantom{0} & 0.014\phantom{0} & 0.009\phantom{0} & \phantom{$<$}0.009\phantom{0} & 0.006\phantom{0} & \phantom{$<$}0.001\phantom{0} \\
$\Kstarbtwo(1430)^0$ $\sqrt{3}S$ & 0.15\phantom{00} & 0.08\phantom{00} & 0.07\phantom{00} & $<$0.01\phantom{00} & 0.02\phantom{00} & $<$0.01\phantom{00} \\
\midrule
$\Lz(1405)$ & 0.28\phantom{00} & 0.08\phantom{00} & 0.05\phantom{00} & \phantom{$<$}0.05\phantom{00} & 0.02\phantom{00} & \phantom{$<$}0.01\phantom{00} \\
$\Lz(1520)$ & 0.11\phantom{00} & 0.06\phantom{00} & 0.04\phantom{00} & \phantom{$<$}0.04\phantom{00} & 0.01\phantom{00} & $<$0.01\phantom{00} \\
$\Lz(1600)$ & 0.39\phantom{00} & 0.08\phantom{00} & 0.07\phantom{00} & \phantom{$<$}0.03\phantom{00} & 0.01\phantom{00} & $<$0.01\phantom{00} \\
$\Lz(1670)$ & 0.13\phantom{00} & 0.11\phantom{00} & 0.10\phantom{00} & \phantom{$<$}0.03\phantom{00} & 0.03\phantom{00} & \phantom{$<$}0.01\phantom{00} \\
$\Lz(1690)$ & 0.07\phantom{00} & 0.12\phantom{00} & 0.12\phantom{00} & \phantom{$<$}0.01\phantom{00} & 0.01\phantom{00} & $<$0.01\phantom{00} \\
$\Lz(1710)$ & 0.35\phantom{00} & 0.06\phantom{00} & 0.05\phantom{00} & \phantom{$<$}0.03\phantom{00} & 0.02\phantom{00} & \phantom{$<$}0.02\phantom{00} \\
$\Lz(1800)$ & 0.95\phantom{00} & 0.71\phantom{00} & 0.68\phantom{00} & \phantom{$<$}0.19\phantom{00} & 0.03\phantom{00} & \phantom{$<$}0.02\phantom{00} \\
$\Lz(1810)$ & 0.43\phantom{00} & 0.08\phantom{00} & 0.07\phantom{00} & \phantom{$<$}0.03\phantom{00} & 0.01\phantom{00} & \phantom{$<$}0.02\phantom{00} \\
$\Lz(1820)$ & 0.19\phantom{00} & 0.21\phantom{00} & 0.21\phantom{00} & \phantom{$<$}0.02\phantom{00} & 0.01\phantom{00} & \phantom{$<$}0.01\phantom{00} \\
$\Lz(1830)$ & 0.91\phantom{00} & 0.41\phantom{00} & 0.41\phantom{00} & $<$0.01\phantom{00} & 0.02\phantom{00} & $<$0.01\phantom{00} \\
$\Lz(1890)$ & 0.51\phantom{00} & 0.21\phantom{00} & 0.20\phantom{00} & \phantom{$<$}0.04\phantom{00} & 0.04\phantom{00} & \phantom{$<$}0.01\phantom{00} \\
$\Lz(2000)$ & 0.13\phantom{00} & 0.05\phantom{00} & 0.05\phantom{00} & $<$0.01\phantom{00} & 0.01\phantom{00} & $<$0.01\phantom{00} \\
\midrule
$\Kstarbzz(700)^0$ & 0.10\phantom{00} & 0.06\phantom{00} & 0.05\phantom{00} & $<$0.01\phantom{00} & 0.02\phantom{00} & $<$0.01\phantom{00} \\
$\Kstarbzz(1430)^0$ & 0.081\phantom{0} & 0.054\phantom{0} & 0.054\phantom{0} & \phantom{$<$}0.010\phantom{0} & 0.003\phantom{0} & $<$0.001\phantom{0} \\
\midrule
$\Deltares(1232)^{++}$ & 0.066\phantom{0} & 0.019\phantom{0} & 0.017\phantom{0} & \phantom{$<$}0.004\phantom{0} & 0.007\phantom{0} & \phantom{$<$}0.001\phantom{0} \\
$\Deltares(1600)^{++}$ & 0.27\phantom{00} & 0.06\phantom{00} & 0.05\phantom{00} & \phantom{$<$}0.03\phantom{00} & 0.01\phantom{00} & $<$0.01\phantom{00} \\
$\Deltares(1620)^{++}$ & 0.31\phantom{00} & 0.22\phantom{00} & 0.21\phantom{00} & \phantom{$<$}0.02\phantom{00} & 0.01\phantom{00} & $<$0.01\phantom{00} \\
$\Deltares(1700)^{++}$ & 0.17\phantom{00} & 0.24\phantom{00} & 0.24\phantom{00} & $<$0.01\phantom{00} & 0.03\phantom{00} & $<$0.01\phantom{00} \\
\bottomrule
\end{tabular}
\end{table}

\clearpage

%% file: Authorship_LHCb-PAPER-2024-034.tex
\centerline
{\large\bf LHCb collaboration}
\begin
{flushleft}
\small
R.~Aaij$^{38}$\lhcborcid{0000-0003-0533-1952},
A.S.W.~Abdelmotteleb$^{57}$\lhcborcid{0000-0001-7905-0542},
C.~Abellan~Beteta$^{51}$\lhcborcid{0009-0009-0869-6798},
F.~Abudin{\'e}n$^{57}$\lhcborcid{0000-0002-6737-3528},
T.~Ackernley$^{61}$\lhcborcid{0000-0002-5951-3498},
A. A. ~Adefisoye$^{69}$\lhcborcid{0000-0003-2448-1550},
B.~Adeva$^{47}$\lhcborcid{0000-0001-9756-3712},
M.~Adinolfi$^{55}$\lhcborcid{0000-0002-1326-1264},
P.~Adlarson$^{82}$\lhcborcid{0000-0001-6280-3851},
C.~Agapopoulou$^{14}$\lhcborcid{0000-0002-2368-0147}
% ...

{\footnotesize \it

% ...
$^{14}$Universit{\'e} Paris-Saclay, CNRS/IN2P3, IJCLab, Orsay, France\\
% ...
$^{38}$Nikhef National Institute for Subatomic Physics, Amsterdam, Netherlands\\
% ...
$^{47}$Instituto Galego de F{\'\i}sica de Altas Enerx{\'\i}as (IGFAE), Universidade de Santiago de Compostela, Santiago de Compostela, Spain\\
% ...
$^{51}$Physik-Institut, Universit{\"a}t Z{\"u}rich, Z{\"u}rich, Switzerland\\
% ...
$^{55}$H.H. Wills Physics Laboratory, University of Bristol, Bristol, United Kingdom\\
% ...
$^{57}$Department of Physics, University of Warwick, Coventry, United Kingdom\\
% ...
$^{60}$School of Physics and Astronomy, University of Glasgow, Glasgow, United Kingdom\\
$^{61}$Oliver Lodge Laboratory, University of Liverpool, Liverpool, United Kingdom\\
% ...
$^{69}$Syracuse University, Syracuse, NY, United States\\
% ...
$^{82}$Department of Physics and Astronomy, Uppsala University, Uppsala, Sweden, associated to $^{60}$\\
% ...
}
\end{flushleft}

%% file: LHCb-DP.bib
@article{LHCb-DP-2018-004,
      author         = "Dominik M{\"u}ller and Marco Clemencic and Gloria Corti and Marco Gersabeck",
      title         = "{ReDecay: A novel approach to speed up the simulation at LHCb}",
      eprint         = "1810.10362",
      archivePrefix  = "arXiv",
      primaryClass   = "hep-ex",
      report         = "LHCb-DP-2018-004",
      year           = "2018",
      journal        = "Eur. Phys. J.",
      volume         = "C78",
      pages          = "1009",
      doi            = "10.1140/epjc/s10052-018-6469-6",
      }

@article{LHCb-DP-2018-001,
      author         = "Aaij, R. and others",
      title          = "{Selection and processing of calibration samples to measure the particle identification performance of the LHCb experiment in Run 2}",
      eprint         = "1803.00824",
      archivePrefix  = "arXiv",
      primaryClass   = "hep-ex",
      report         = "LHCb-DP-2018-001",
      year           = "2019",
      journal        = "Eur. Phys. J. Tech. Instr.",
      volume         = "6",
      pages          = "1",
      doi            = "10.1140/epjti/s40485-019-0050-z",
}

@article{LHCb-DP-2014-002,
      author         = "Aaij, R. and others",
      title          = "{LHCb detector performance}",
      collaboration  = "LHCb collaboration",
      journal        = "Int. J. Mod. Phys.",
      volume         = "A30",
      pages          = "1530022",
      doi            = "10.1142/S0217751X15300227",
      year           = "2015",
      eprint         = "1412.6352",
      archivePrefix  = "arXiv",
      primaryClass   = "hep-ex",
      report         = "LHCB-DP-2014-002, CERN-PH-EP-2014-290",
}

@article{LHCb-DP-2008-001,
      author         = "Alves~Jr., A. A. and others",
      title          = "{The \lhcb detector at the LHC}",
      collaboration  = "LHCb collaboration",
      journal        = "JINST",
      volume         = "3",
      pages          = "S08005",
      doi            = "10.1088/1748-0221/3/08/S08005",
      year           = "2008",
      number         = "LHCb-DP-2008-001",
}


%% file: LHCb-PAPER.bib
@article{LHCb-PAPER-2024-043,
      author         = "Aaij, R. and others",
      title          = "{Study of $\Lambda^{0}_{b}$ and $\Xi^{0}_{b}$ decays to $\Lambda h^+ h^{\prime -}$ and evidence for \CP violation in $\Lambda^{0}_{b} \to \Lambda K^+ K^-$}",
      collaboration  = "LHCb collaboration",
      report         = "{LHCb-PAPER-2024-043, CERN-EP-2024-281}",
      eprint         = "2411.15441",
      archivePrefix  = "arXiv",
      primaryClass   = "hep-ex",
      journal        = "{Phys. Rev. Lett.}",
      volume       = "134",
      pages        = "101802",
      year           = "2025",
      doi             = "10.1103/PhysRevLett.134.101802"
}

@article{LHCb-PAPER-2022-002,
      author         = "Aaij, R. and others",
      title          = "{Amplitude analysis of the $\Lc \to \proton \Km\pip$ decay and $\Lc$ baryon polarization measurement in semileptonic beauty hadron decays}",
      collaboration  = "LHCb collaboration",
      report         = "{LHCb-PAPER-2022-002, CERN-EP-2022-124}",
      eprint         = "2208.03262",
      archivePrefix  = "arXiv",
      primaryClass   = "hep-ex",
      year           = "2023",
      journal        = "Phys. Rev.",
      volume         = "D108",
      pages          = "012023",
      doi            = "10.1103/PhysRevD.108.012023",
}

@article{LHCb-PAPER-2019-028,
      author         = "Aaij, R. and others",
      title          = "{Search for \CP violation and observation of $P$ violation in \decay{\Lb}{p\pim\pip\pim} decays}",
      collaboration  = "LHCb collaboration",
      report         = "{LHCb-PAPER-2019-028 CERN-EP-2019-256}",
      eprint         = "1912.10741",
      archivePrefix  = "arXiv",
      primaryClass   = "hep-ex",
      year           = "2020",
      journal        = "Phys. Rev.",
      volume         = "D102",
      pages          = "051101",
      doi            = "10.1103/PhysRevD.102.051101",
}

@article{LHCb-PAPER-2019-026,
      author         = "Aaij, R. and others",
      title          = "{Search for \CP violation in \mbox{\decay{\Xicp}{p\Km\pip}} decays with model-independent techniques}",
      collaboration  = "LHCb collaboration",
      report         = "{LHCb-PAPER-2019-026, CERN-EP-2020-069}",
      eprint         = "2006.03145",
      archivePrefix  = "arXiv",
      primaryClass   = "hep-ex",
      year           = "2020",
      journal        = "Eur. Phys. J.",
      volume         = "C80",
      pages          = "986",
      doi            = "10.1140/epjc/s10052-020-8365-0",

}

@article{LHCb-PAPER-2018-044,
      author         = "Aaij, R. and others",
      title          = "{Measurement of \CP asymmetries in charmless four-body \Lb and \Xibz decays}",
      collaboration  = "LHCb collaboration",
      report         = "{LHCb-PAPER-2018-044 CERN-EP-2019-013}",
      eprint         = "1903.06792",
      archivePrefix  = "arXiv",
      primaryClass   = "hep-ex",
      year           = "2019",
      journal        = "Eur. Phys. J.",
      volume         = "C79",
      pages          = "745",
      doi            = "10.1140/epjc/s10052-019-7218-1",
}

@article{LHCb-PAPER-2018-025,
      author         = "Aaij, R. and others",
      title          = "{Search for \CP violation in \mbox{\decay{\Lb}{p\Km} and \mbox{\decay{\Lb}{p\pim}}} decays}",
      collaboration  = "LHCb collaboration",
      report         = "{LHCb-PAPER-2018-025 CERN-EP-2018-189}",
      eprint         = "1807.06544",
      archivePrefix  = "arXiv",
      primaryClass   = "hep-ex",
      year           = "2018",
      journal        = "Phys. Lett.",
      volume         = "B784",
      pages          = "101",
      doi            = "10.1016/j.physletb.2018.10.039",
}

@article{LHCb-PAPER-2018-001,
      author         = "Aaij, R. and others",
      title          = "{Search for \CP violation using triple product asymmetries in \mbox{\decay{\Lb}{p\Km\pip\pim}}, \mbox{\decay{\Lb}{p\Km\Kp\Km}}, and \mbox{\decay{\Xires^0_b}{p\Km\Km\pip}} decays}",
      collaboration  = "LHCb collaboration",
      report         = "{LHCb-PAPER-2018-001 CERN-EP-2018-081}",
      eprint         = "1805.03941",
      archivePrefix  = "arXiv",
      primaryClass   = "hep-ex",
      year           = "2018",
      journal        = "JHEP",
      volume         = "08",
      pages          = "039",
      doi            = "10.1007/JHEP08(2018)039",
}

@article{LHCb-PAPER-2017-044,
      author         = "Aaij, R. and others",
      title          = "{Search for \CP violation in \mbox{\decay{\Lc}{p \Km \Kp}} and \mbox{\decay{\Lc}{p\pim\pip}} decays}",
      collaboration  = "LHCb collaboration",
      report         = "{LHCb-PAPER-2017-044 CERN-EP-2017-316}",
      eprint         = "1712.07051",
      archivePrefix  = "arXiv",
      primaryClass   = "hep-ex",
      year           = "2018",
      journal        = "JHEP",
      volume         = "03",
      pages          = "182",
      doi            = "10.1007/JHEP03(2018)182",
}

@article{LHCb-PAPER-2016-059,
      author         = "Aaij, R. and others",
      title          = "{Observation of the decay \mbox{\decay{\Lb}{\proton\Km\mumu}} and search for \CP violation}",
      collaboration  = "LHCb collaboration",
      year           = "2017",
      journal        = "JHEP",
      volume         = "06",
      pages          = "108",
      doi            = "10.1007/JHEP06(2017)108",
      report         = "{LHCb-PAPER-2016-059 CERN-EP-2017-032}",
      eprint         = "1703.00256",
      archivePrefix  = "arXiv",
      primaryClass   = "hep-ex",
}

@article{LHCb-PAPER-2013-011,
      author         = "Aaij, R. and others",
      title          = "{Precision measurement of \D meson mass differences}",
      collaboration  = "LHCb collaboration",
      journal        = "JHEP",
      volume         = "06",
      pages          = "065",
      doi            = "10.1007/JHEP06(2013)065",
      year           = "2013",
      eprint         = "1304.6865",
      archivePrefix  = "arXiv",
      primaryClass   = "hep-ex",
      report         = "CERN-PH-EP-2013-053 LHCb-PAPER-2013-011",
}

@article{LHCb-PAPER-2012-048,
      author         = "Aaij, R. and others",
      title          = "{Measurements of the \Lb, \Xibm, and \Omegab baryon masses}",
      collaboration  = "LHCb collaboration",
      journal        = "Phys. Rev. Lett.",
      volume         = "110",
      pages          = "182001",
      doi            = "10.1103/PhysRevLett.110.182001",
      year           = "2013",
      eprint         = "1302.1072",
      archivePrefix  = "arXiv",
      primaryClass   = "hep-ex",
      report         = "CERN-PH-EP-2013-013 LHCb-PAPER-2012-048",
}


%% file: main.bib
@article{Botella:2016ksl,
      author         = "Botella, F. J. and Garcia Martin, L. M. and Marangotto,
                        D. and Vidal, F. Martinez and Merli, A. and Neri, N. and
                        Oyanguren, A. and Vidal, J. Ruiz",
      title          = "{On the search for the electric dipole moment of strange
                        and charm baryons at LHC}",
      journal        = "Eur. Phys. J.",
      volume         = "C77",
      year           = "2017",
      number         = "3",
      pages          = "181",
      doi            = "10.1140/epjc/s10052-017-4679-y",
      eprint         = "1612.06769",
      archivePrefix  = "arXiv",
      primaryClass   = "hep-ex",
      SLACcitation   = "%%CITATION = ARXIV:1612.06769;%%"
}

@article{Bagli:2017foe,
      author         = "Bagli, E. and others",
      title          = "{Electromagnetic dipole moments of charged baryons with
                        bent crystals at the LHC}",
      journal        = "Eur. Phys. J.",
      volume         = "C77",
      year           = "2017",
      number         = "12",
      pages          = "828",
      doi            = "10.1140/epjc/s10052-017-5400-x",
      eprint         = "1708.08483",
      archivePrefix  = "arXiv",
      primaryClass   = "hep-ex",
      SLACcitation   = "%%CITATION = ARXIV:1708.08483;%%"
}

@article{JacobWick,
title = "On the general theory of collisions for particles with spin",
journal = "Annals of Physics",
volume = "7",
number = "4",
pages = "404 - 428",
year = "1959",
issn = "0003-4916",
doi = "https://doi.org/10.1016/0003-4916(59)90051-X",
url = "http://www.sciencedirect.com/science/article/pii/000349165990051X",
author = "Jacob, M. and Wick, G.C."
}

@article{Flatte:1976xu,
      author         = "Flatt\'e, Stanley M.",
      title          = "{Coupled-channel analysis of the $\pi \eta$ and $K \bar K$
                        systems near $K \bar K$ threshold}",
      journal        = "Phys. Lett.",
      volume         = "B63",
      year           = "1976",
      pages          = "224-227",
      doi            = "10.1016/0370-2693(76)90654-7",
      reportNumber   = "CERN-EP-PHYS-76-8",
      SLACcitation   = "%%CITATION = PHLTA,63B,224;%%"
}

@article{Galanti:2015pqa,
      author         = "Galanti, Mario and Giammanco, Andrea and Grossman, Yuval
                        and Kats, Yevgeny and Stamou, Emmanuel and Zupan, Jure",
      title          = "{Heavy baryons as polarimeters at colliders}",
      journal        = "JHEP",
      volume         = "11",
      year           = "2015",
      pages          = "067",
      doi            = "10.1007/JHEP11(2015)067",
      eprint         = "1505.02771",
      archivePrefix  = "arXiv",
      primaryClass   = "hep-ph",
      reportNumber   = "CP3-15-12",
      SLACcitation   = "%%CITATION = ARXIV:1505.02771;%%"
}

@article{Mannel:1991bs,
      author         = "Mannel, Thomas and Schuler, Gerhard A.",
      title          = "{Semileptonic decays of bottom baryons at LEP}",
      journal        = "Phys. Lett.",
      volume         = "B279",
      year           = "1992",
      pages          = "194-200",
      doi            = "10.1016/0370-2693(92)91864-6",
      reportNumber   = "DESY-91-095",
      SLACcitation   = "%%CITATION = PHLTA,B279,194;%%"
}

@article{Falk:1993rf,
      author         = "Falk, Adam F. and Peskin, Michael E.",
      title          = "{Production, decay, and polarization of excited heavy
                        hadrons}",
      journal        = "Phys. Rev.",
      volume         = "D49",
      year           = "1994",
      pages          = "3320-3332",
      doi            = "10.1103/PhysRevD.49.3320",
      eprint         = "hep-ph/9308241",
      archivePrefix  = "arXiv",
      primaryClass   = "hep-ph",
      reportNumber   = "SLAC-PUB-6311, JHU-TIPAC-930019",
      SLACcitation   = "%%CITATION = HEP-PH/9308241;%%"
}

@article{Brun:1997pa,
      author         = "Brun, R. and Rademakers, F.",
      title          = "{ROOT: An object oriented data analysis framework}",
      booktitle      = "{New computing techniques in physics research V.
                        Proceedings, 5th International Workshop, AIHENP '96,
                        Lausanne, Switzerland, September 2-6, 1996}",
      journal        = "Nucl. Instrum. Meth.",
      volume         = "A389",
      year           = "1997",
      pages          = "81-86",
      doi            = "10.1016/S0168-9002(97)00048-X",
      SLACcitation   = "%%CITATION = NUIMA,A389,81;%%"
}

@misc{TFA,
title={ {TensorFlowAnalysis}: A collection of useful functions and example scripts for performing amplitude fits using {TensorFlow}},
note={\url{https://gitlab.cern.ch/poluekt/TensorFlowAnalysis}},
}

@article{Bugg:2005xx,
    author = "Bugg, D. V.",
    title = "{The kappa in E791 data for \decay{D}{\PK \pi \pi}}",
    eprint = "hep-ex/0510019",
    archivePrefix = "arXiv",
    doi = "10.1016/j.physletb.2005.11.019",
    journal = "Phys. Lett.",
    volume = "B632",
    pages = "471--474",
    year = "2006"
}

@article{James:1975dr,
      author         = "James, F. and Roos, M.",
      title          = "{Minuit: A system for function minimization and analysis
                        of the parameter errors and correlations}",
      journal        = "Comput. Phys. Commun.",
      volume         = "10",
      year           = "1975",
      pages          = "343-367",
      doi            = "10.1016/0010-4655(75)90039-9",
      reportNumber   = "CERN-DD-75-20",
      SLACcitation   = "%%CITATION = CPHCB,10,343;%%"
}

@article{Marangotto:2019ucc,
      author         = "Marangotto, Daniele",
      title          = "{Helicity amplitudes for generic multibody particle decays featuring multiple decay chains}",
      year           = "2020",
      eprint         = "1911.10025",
      archivePrefix  = "arXiv",
      primaryClass   = "hep-ph",
      doi = "10.1155/2020/6674595",
      journal = "Adv. High Energy Phys.",
      volume = "2020",
      pages = "6674595",
      SLACcitation   = "%%CITATION = ARXIV:1911.10025;%%"
}

@thesis{Marangotto:2713231,
      author        = "Marangotto, Daniele",
      title         = "{Amplitude analysis and polarisation measurement of the
                       $\Lc$ baryon in $\proton\Km\pip$ final state for
                       electromagnetic dipole moment experiment}",
      reportNumber  = "CERN-THESIS-2020-015",
      url           = "https://cds.cern.ch/record/2713231",
      note          = "PhD thesis, Universit\`a degli studi di Milano, Presented 16 Mar 2020, \url{https://cds.cern.ch/record/2713231}",
}

@article{Marangotto:2020ead,
    author = "Marangotto, Daniele",
    title = "{Extracting maximum information from polarised baryon decays via amplitude analysis: The $\Lc \to \proton\Km\pip$ case}",
    eprint = "2004.12318",
    archivePrefix = "arXiv",
    primaryClass = "hep-ph",
    doi = "10.1155/2020/7463073",
    journal = "Adv. High Energy Phys.",
    volume = "2020",
    pages = "7463073",
    year = "2020"
}

@article{Aiola:2020yam,
    author = "Aiola, S. and others",
    title = "{Progress towards the first measurement of charm baryon dipole moments}",
    eprint = "2010.11902",
    archivePrefix = "arXiv",
    primaryClass = "hep-ex",
    doi = "10.1103/PhysRevD.103.072003",
    journal = "Phys. Rev.",
    volume = "D103",
    pages = "072003",
    year = "2020"
}

@article{Davier:1992nw,
      author         = "Davier, M. and Duflot, L. and Le Diberder, F. and Rouge,
                        A.",
      title          = "{The optimal method for the measurement of tau
                        polarization}",
      journal        = "Phys. Lett.",
      volume         = "B306",
      year           = "1993",
      pages          = "411-417",
      doi            = "10.1016/0370-2693(93)90101-M",
      reportNumber   = "LAL-92-73, X-LPNHE-92-22",
      SLACcitation   = "%%CITATION = PHLTA,B306,411;%%"
}

@article{Poluektov:2014rxa,
    author = "Poluektov, Anton",
    title = "{Kernel density estimation of a multidimensional efficiency profile}",
    eprint = "1411.5528",
    archivePrefix = "arXiv",
    primaryClass = "physics.data-an",
    doi = "10.1088/1748-0221/10/02/P02011",
    journal = "JINST",
    volume = "10",
    number = "02",
    pages = "P02011",
    year = "2015"
}

@article{Belle:2022cbs,
    author = "Yang, S. B. and others",
    collaboration = "Belle",
    title = "{Observation of a threshold cusp at the $\Lz\Peta$ threshold in the $\proton\Km$ mass spectrum with $\Lc \to \proton\Km\pip$ decays}",
    eprint = "2209.00050",
    archivePrefix = "arXiv",
    primaryClass = "hep-ex",
    reportNumber = "Belle Preprint 2022-21, KEK Preprint 2022-29",
    doi = "10.1103/PhysRevD.108.L031104",
    journal = "Phys. Rev.",
    volume = "D108",
    number = "3",
    pages = "L031104",
    year = "2023"
}

@article{Zhang:2019xdm,
    author = "Zhang, Jiao and An, Xiuyun and Sun, Ruirui and Su, Jianfeng",
    title = "{Probing new physics in semileptonic $\Xib\to \Lz (\Xic)\taum\neutb$ decays}",
    doi = "10.1140/epjc/s10052-019-7373-4",
    journal = "Eur. Phys. J.",
    volume = "C79",
    number = "10",
    pages = "863",
    year = "2019"
}

@article{Wang:2021ydv,
    author = "Wang, Shuai-Wei",
    title = "{$\Xib \to \Xic \Ptau\neutb$ decay in new physics models}",
    doi = "10.1007/s10773-021-04721-3",
    journal = "Int. J. Theor. Phys.",
    volume = "60",
    number = "3",
    pages = "982--993",
    year = "2021"
}

@article{Zhang:2019jax,
    author = "Zhang, Jiao and Su, Jianfeng and Zeng, Qingguo",
    title = "{Contributions of vector leptoquark to $\Xib \to \Xic \Ptau\neutb$ decay}",
    doi = "10.1016/j.nuclphysb.2018.11.006",
    journal = "Nucl. Phys.",
    volume = "B938",
    pages = "131--142",
    year = "2019"
}

@article{Fomin:2017ltw,
    author = "Fomin, A. S. and others",
    title = "{Feasibility of measuring the magnetic dipole moments of the charm baryons at the LHC using bent crystals}",
    eprint = "1705.03382",
    archivePrefix = "arXiv",
    primaryClass = "hep-ph",
    doi = "10.1007/JHEP08(2017)120",
    journal = "JHEP",
    volume = "08",
    pages = "120",
    year = "2017"
}

@article{Fomin:2019wuw,
    author = "Fomin, A. S. and Barsuk, S. and Korchin, A. Yu. and Kovalchuk, V. A. and Kou, E. and Liul, M. and Natochii, A. and Niel, E. and Robbe, P. and Stocchi, A.",
    title = "{The prospect of charm quark magnetic moment determination}",
    eprint = "1909.04654",
    archivePrefix = "arXiv",
    primaryClass = "hep-ph",
    doi = "10.1140/epjc/s10052-020-7891-0",
    journal = "Eur. Phys. J. C",
    volume = "80",
    number = "5",
    pages = "358",
    year = "2020"
}

@article{Mirarchi:2019vqi,
    author = "Mirarchi, D. and Fomin, A. S. and Redaelli, S. and Scandale, W.",
    title = "{Layouts for fixed-target experiments and dipole moment measurements of short-lived baryons using bent crystals at the LHC}",
    eprint = "1906.08551",
    archivePrefix = "arXiv",
    primaryClass = "physics.acc-ph",
    doi = "10.1140/epjc/s10052-020-08466-x",
    journal = "Eur. Phys. J. C",
    volume = "80",
    number = "10",
    pages = "929",
    year = "2020"
}


%% file: standard.bib
@article{PDG2024,
     author    = "Navas, S. and others",
    collaboration = "Particle Data Group",
     title     = "{\href{http://pdg.lbl.gov/}{Review of particle physics}}",
     journal   = "Phys. Rev",
     year = {2024},
     number = {8},
     volume      = "D110",
     pages     = "030001",
     doi = "10.1103/PhysRevD.110.030001"
}

@article{LHCb-PROC-2015-018,
      author        = "Likhomanenko, T. and others",
      title         = "{LHCb topological trigger reoptimization}",
      journal       = "J. Phys. Conf. Ser.",
      volume        = "664",
      pages         = "082025",
      doi           = "10.1088/1742-6596/664/8/082025",
      month         = "Oct",
      year          = "2015",
      reportNumber  = "LHCb-PROC-2015-018",
      eprint        = "1510.00572",
      archivePrefix = "arXiv"
}

@article{LHCb-PROC-2011-006,
  author="Clemencic, M and others",
  title="{The \lhcb simulation application, Gauss: Design, evolution and experience}",
  journal="J. Phys. Conf. Ser.",
  volume={331},
  pages={032023},
  doi={10.1088/1742-6596/331/3/032023},
  year={2011},
}

@article{LHCb-PROC-2010-056,
      author         = "Belyaev, I. and others",
      title          = "{Handling of the generation of primary events
                         in Gauss, the LHCb simulation framework}",
      journal="J. Phys. Conf. Ser.",
      volume={331},
      pages={032047},
      doi={10.1088/1742-6596/331/3/032047},
      year={2011},
}

@article{Sjostrand:2006za,
      author         = {Sj\"{o}strand, Torbj\"{o}rn and Mrenna, Stephen and
                       Skands, Peter"},
      title          = "{PYTHIA 6.4 physics and manual}",
      journal        = "JHEP",
      volume         = "05",
      pages          = "026",
      doi            = "10.1088/1126-6708/2006/05/026",
      year           = "2006",
      eprint         = "hep-ph/0603175",
      archivePrefix  = "arXiv",
      primaryClass   = "hep-ph",
}

@article{Sjostrand:2007gs,
      author         = {Sj\"{o}strand, Torbj\"{o}rn and Mrenna, Stephen and
                        Skands, Peter"},
      title          = "{A brief introduction to PYTHIA 8.1}",
      journal        = "Comput. Phys. Commun.",
      volume         = "178",
      pages          = "852-867",
      doi            = "10.1016/j.cpc.2008.01.036",
      year           = "2008",
      eprint         = "0710.3820",
      archivePrefix  = "arXiv",
      primaryClass   = "hep-ph",
      reportNumber   = "CERN-LCGAPP-2007-04, LU-TP-07-28,
                        FERMILAB-PUB-07-512-CD-T",
}

@Article{Agostinelli:2002hh,
     author    = "Agostinelli, S. and others",
 collaboration = "Geant4 collaboration",
     title     = "{Geant4: A simulation toolkit}",
     journal   = "Nucl. Instrum. Meth.",
     volume    = "A506",
     year      = "2003",
     pages     = "250",
     doi       = "10.1016/S0168-9002(03)01368-8",
}

@article{Allison:2006ve,
      author         = "Allison, John and Amako, K. and Apostolakis, J. and
                        Araujo, H. and Dubois, P.A. and others",
 collaboration = "Geant4 collaboration",
      title          = "{Geant4 developments and applications}",
      journal        = "IEEE Trans.Nucl.Sci.",
      volume         = "53",
      pages          = "270",
      doi            = "10.1109/TNS.2006.869826",
      year           = "2006",
      reportNumber   = "SLAC-PUB-11870",
}

@Article{Lange:2001uf,
     author    = "Lange, D. J.",
     title     = "{The EvtGen particle decay simulation package}",
     journal   = "Nucl. Instrum. Meth.",
     volume    = "A462",
     year      = "2001",
     pages     = "152-155",
     doi       = "10.1016/S0168-9002(01)00089-4",
}

@article{davidson2015photos,
	author="Davidson, N. and Przedzinski, T. and Was, Z.",
	title = "{PHOTOS interface in C++: Technical and physics documentation}",
      	eprint={1011.0937},
      	archivePrefix={arXiv},
      	primaryClass={hep-ph},
	journal = {Comp. Phys. Comm.},
	volume = {199},
	pages = {86},
	year = {2016},
	doi = {https://doi.org/10.1016/j.cpc.2015.09.013},
}

@article{BBDT,
      author         = "Gligorov, V. V. and Williams, M.",
      title          = "{Efficient, reliable and fast high-level triggering using a bonsai boosted decision tree}",
      journal        = "JINST",
      volume         = "8",
      pages          = "P02013",
      doi            = "10.1088/1748-0221/8/02/P02013",
      year           = "2013",
      eprint         = "1210.6861",
      archivePrefix  = "arXiv",
      primaryClass   = "physics.ins-det",

}

@article{Pivk:2004ty,
      author         = "Pivk, Muriel and Le Diberder, Francois R.",
      title          = "{sPlot: A statistical tool to unfold data distributions}",
      journal        = "Nucl. Instrum. Meth.",
      volume         = "A555",
      pages          = "356-369",
      doi            = "10.1016/j.nima.2005.08.106",
      year           = "2005",
      eprint         = "physics/0402083",
      archivePrefix  = "arXiv",
      primaryClass   = "physics.data-an",
      reportNumber   = "LAL-04-07",
}

@misc{tensorflow2015-whitepaper,
title={ {TensorFlow}: Large-Scale Machine Learning on Heterogeneous Systems},
url={https://www.tensorflow.org/},
note={Software available from tensorflow.org},
author={
    Mart\'{i}n~Abadi and
    Ashish~Agarwal and
    Paul~Barham and
    Eugene~Brevdo and
    Zhifeng~Chen and
    Craig~Citro and
    Greg~S.~Corrado and
    Andy~Davis and
    Jeffrey~Dean and
    Matthieu~Devin and
    Sanjay~Ghemawat and
    Ian~Goodfellow and
    Andrew~Harp and
    Geoffrey~Irving and
    Michael~Isard and
    Yangqing Jia and
    Rafal~Jozefowicz and
    Lukasz~Kaiser and
    Manjunath~Kudlur and
    Josh~Levenberg and
    Dandelion~Man\'{e} and
    Rajat~Monga and
    Sherry~Moore and
    Derek~Murray and
    Chris~Olah and
    Mike~Schuster and
    Jonathon~Shlens and
    Benoit~Steiner and
    Ilya~Sutskever and
    Kunal~Talwar and
    Paul~Tucker and
    Vincent~Vanhoucke and
    Vijay~Vasudevan and
    Fernanda~Vi\'{e}gas and
    Oriol~Vinyals and
    Pete~Warden and
    Martin~Wattenberg and
    Martin~Wicke and
    Yuan~Yu and
    Xiaoqiang~Zheng},
  year={2015},
}

@Misc{mciteplus,
  author = 	 {Shell, Michael},
  title = 	 {Mciteplus: Enhanced multicitations},
  howpublished = {\href{http://www.michaelshell.org/tex/mciteplus/} {http://www.michaelshell.org/tex/mciteplus/}},
}

@article{Rogozhnikov:2016bdp,
      author         = "Rogozhnikov, A.",
      title          = "{Reweighting with boosted decision trees}",
      booktitle      = "{Proceedings, 17th International Workshop on Advanced
                        Computing and Analysis Techniques in Physics Research
                        (ACAT 2016): Valparaiso, Chile, January 18-22, 2016}",
      journal        = "J. Phys. Conf. Ser.",
      volume         = "762",
      year           = "2016",
      number         = "1",
      pages          = "012036",
      doi            = "10.1088/1742-6596/762/1/012036",
      eprint         = "1608.05806",
      archivePrefix  = "arXiv",
      primaryClass   = "physics.data-an",
      note           = "\url{https://github.com/arogozhnikov/hep_ml}",
}

@book{Breiman,
  author = 	 {Breiman, L. and Friedman, J. H. and Olshen,
                  R. A. and Stone, C. J.},
  title = 	 {Classification and regression trees},
  publisher = 	 {Wadsworth international group},
  year = 	 {1984},
  address = 	 {Belmont, California, USA},
}
